\documentstyle[emulateapj,danonecolfloat]{article}
\input psfig.sty

\def\bi#1{\hbox{\boldmath{$#1$}}}
\def\spose#1{\hbox to 0pt{#1\hss}}
\def\simlt{\mathrel{\spose{\lower 3pt\hbox{$\mathchar"218$}}
     \raise 2.0pt\hbox{$\mathchar"13C$}}}
\def\simgt{\mathrel{\spose{\lower 3pt\hbox{$\mathchar"218$}}
     \raise 2.0pt\hbox{$\mathchar"13E$}}}
\def\simpropto{\mathrel{\spose{\lower 3pt\hbox{$\mathchar"218$}}
     \raise 2.0pt\hbox{$\propto$}}}
\newcommand\lsim{\mathrel{\rlap{\lower4pt\hbox{\hskip1pt$\sim$}}
        \raise1pt\hbox{$<$}}}
\newcommand\gsim{\mathrel{\rlap{\lower4pt\hbox{\hskip1pt$\sim$}}
        \raise1pt\hbox{$>$}}}
\def\bi#1{\hbox{\boldmath{$#1$}}}    

\begin{document}
\twocolumn[



\submitted{May 29, 2002}

\title{Intrinsic/Extrinsic Density-Ellipticity Correlations and Galaxy-Galaxy Lensing}
\author{Lam Hui and Jun Zhang}
\affil{Department of Physics, Columbia University\\538, West 120th Street,
New York, NY 10027\\ {\tt lhui@astro.columbia.edu, jz203@columbia.edu}}
                                          
\begin{abstract}
We compute both extrinsic (lensing) and intrinsic contributions to
the (galaxy-)density-ellipticity correlation function, the latter
done using current analytic theories of tidal alignment. 
The gravitational lensing contribution has two components:
one is analogous to galaxy-galaxy lensing and the other arises from magnification bias 
-- that gravitational lensing induces a modulation of the galaxy density as well 
as ellipticity. On the other hand, the intrinsic alignment contribution vanishes, even
after taking into account source clustering corrections, which
suggests the density-ellipticity correlation might be an interesting
diagnostic in differentiating between intrinsic and extrinsic alignments.
{\it However}, an important assumption, commonly adopted by current
analytic alignment theories, is the Gaussianity of the tidal field.
Inevitable non-Gaussian fluctuations from gravitational instability induces a non-zero
intrinsic density-ellipticity correlation, which we estimate.
We also argue that non-Gaussian contributions to the intrinsic 
{\it ellipticity-ellipticity} correlation are often non-negligible. This leads to
a linear rather than, as is commonly assumed, 
quadratic scaling with the power spectrum on sufficiently large scales. 
Finally, we estimate the contribution of intrinsic alignment to low redshift
galaxy-galaxy lensing measurements (e.g. SDSS), due to the partial overlap between
foreground and background galaxies: the intrinsic contamination is about
$10 - 30\%$ at $10 '$. Uncertainties in this estimate are discussed.
\end{abstract}


]

\section{Introduction}
\label{intro}

Weak gravitational lensing provides a direct probe of the large scale
distribution of mass (Gunn 1967,Miralda-Escude 1991, Blandford et al. 1991, Kaiser 1992; 
see Mellier 1999 and Bartelmann \& Schneider 2001 for reviews). 
Analyses by several groups have demonstrated that measurements
of galaxy shapes can be made sufficiently precise to detect such an effect 
(Wittman et al. 2000, van Waerbeke et al.  2000, Bacon et al. 2000, Kaiser et al. 2000, 
Maoli et al. 2001, Rhodes et al. 2001). 
As the precision improves, it is important to 
understand possible small contaminations to the lensing signal. 
An assumption that has generally been made is that there is no intrinsic
correlation between shapes/orientations of separate galaxies, 
other than through gravitational lensing (extrinsic alignment). 
Recently, it has been pointed out that some amount of intrinsic alignment
might be expected due to long range tidal correlations (Lee \& Pen 2000, 
Pen, Lee \& Seljak 2000, 
Croft \& Metzler 2000, Heavens et al. 2000,  Brown et al. 2000, Catelan et al. 2001
Crittenden et al. 2001, 
Mackey et al. 2001, Porciani et al. 2001, Vitvitska et al. 2001, Maller et al. 2001, van den Bosch et al. 2002
). The estimated level of contamination ranges from weak ($\lsim 10 \%$) for
source galaxies at a median redshift of $1$, to dominant for galaxies at
a redshift of $0.1$ or smaller.

There are several ways one might be able to differentiate intrinsic from
extrinsic correlations. This is important not only from the point of
view of isolating the lensing signal; the intrinsic correlation signal
is also interesting in its own right -- unraveling its origin may teach
us much about the origin of angular momentum of galaxies. 
First of all, if a sufficient number of galaxies with accurate redshifts
(say using photometric redshifts) is present, one can investigate the
scaling of the measured correlations with redshift -- the intrinsic and
extrinsic signals are expected to vary differently with the median redshift
of the source galaxies, as well as the width of the galaxy distribution.
At the moment, such an approach might be challenging to implement due to 
the lack of a large number of accurate redshifts in typical deep lensing surveys.
Second, as is well-known, gravitational lensing introduces an ellipticity-ellipticity
correlation that is curl-free (e.g. Stebbins 1996)
\footnote{Schneider et al. (2001) recently pointed out there is a non-zero 
curl contribution (B-mode) from lensing due to source redshift clustering. 
The signal is only significant, however, at small angular scales, below
about an arcminute or so.}
, while as emphasized by Pen et al. (2000) and
Crittenden et al. (2000), intrinsic correlations generally carry both a curl
and a curl-free component (or a magnetic and an electric
part). Pen, van Waerbeke \& Mellier (2001) (see also Hoekstra, Yee
\& Gladders 2002) presented some recent
measurements. However, the curl part
of the measured ellipticity correlations is also often used as a diagnostic
for systematic errors such as those introduced by correlated point-spread-function variations.
It would therefore be useful to study other statistics that might help 
separate the intrinsic from the extrinsic signals. 

This was what motivated us to study the density-ellipticity cross correlation.
Most studies so far have focused on ellipticity-ellipticity correlations
$\langle \epsilon_i (\bi{\theta}) \epsilon_j (\bi{\theta'}) \rangle$, where $i$ and $j$
stands for the two components of the ellipticity, and $\bi{\theta}$ and
$\bi{\theta'}$ denote two angular positions in the sky. What we would like to 
investigate in this paper is the density-ellipticity correlation
$\langle \delta_g (\bi{\theta}) \epsilon_i (\bi{\theta'}) \rangle$, where 
$\delta_g (\bi{\theta})$ is the galaxy overdensity 
($\delta_g \equiv (n_g - \bar n_g)/\bar n_g$ where $n_g$ is the galaxy number
density, and $\bar n_g$ is its mean) at angular position $\bi{\theta}$. 
\footnote{We will also examine a correction to this quantity due to
source clustering: $\langle \delta_g (\bi{\theta}) \epsilon_i^\Delta (\bi{\theta'}) \rangle $;
its precise definition will be given in \S \ref{setstage}.}
This quantity is very much analogous to what is often measured in the context
of galaxy-galaxy lensing (Brainerd et al. 1996, Schneider 1998, Fischer et al. 2000). 
The difference is that while in galaxy-galaxy lensing $\delta_g$ is measured
from a foreground population of galaxies and $\epsilon_i$ from a background population,
{\it here, we are mainly interested in $\delta_g$ and $\epsilon_i$ measured from
the same population of galaxies} (although we will have the occasion to study
galaxy-galaxy lensing as well). 

What makes the density-ellipticity correlation, 
$\langle \delta_g \epsilon_i \rangle$,
particularly interesting is the fact that intrinsic alignment makes no 
contributions to it, at least according to current tidal theories of intrinsic 
alignment (Crittenden et al. 2001 [CNPT hereafter] 
Mackey et al. 2001 [MWK hereafter]). This is easy to see given 
that $\delta_g \propto \nabla^2 \phi$ (where $\phi$ is the gravitational potential, and
linear biasing is assumed, which should hold on large scales, where current
tidal theories are supposed to work), and
the intrinsic ellipticity depends quadratically on $\phi$. 
Gaussianity of the gravitational potential fluctuation, assumed by both CNPT and MWK,
guarantees that the expectation value of the resulting cubic product of $\phi$ from
$\langle \delta_g \epsilon_i \rangle$ vanishes. What is less obvious is
whether this continues to hold if the source clustering correction is included
(which we label as $\langle \delta_g \epsilon_i^\Delta \rangle$, and scales
quartically with $\phi$). We will demonstrate in \S \ref{intrinsic} that it does.

In \S \ref{extrinsic}, we work out the gravitational lensing (extrinsic) contribution
to the density-ellipticity correlation. It turns out to have two components, one
very much analogous to galaxy-galaxy lensing, and the other arising from
magnification bias. The relative importance of the two depends on the redshift
distribution of the galaxies, which we illustrate with a few examples.

It appears then that the density-ellipticity correlation might provide an
interesting test of the lensing hypothesis in
ellipticity-ellipticity measurements, particularly in view of the fact
that intrinsic alignment creates no density-ellipticity correlation
according to current tidal theories. However, we expect the latter to
break down when one takes into account inevitable non-Gaussian fluctuations
introduced by gravitational instability.
This is discussed in \S \ref{nonGauss}. In fact, we will argue that 
the existing calculations of the intrinsic {\it ellipticity-ellipticity} correlation, 
by ignoring non-Gaussian contributions, miss a term that
scales linearly with the power spectrum, and that therefore dominates over the Gaussian
(quadratic) term {\it on large scales}.
A simple but crude way to understand the origin of this result is encapsulated in eq. (\ref{crude})
in \S \ref{digression}.

In \S \ref{galgal}, we apply the non-Gaussian calculation to estimate
the contamination from intrinsic alignment in galaxy-galaxy lensing measurements.
As we have discussed, galaxy-galaxy lensing measures the same quantity
$\langle \delta_g \epsilon_i \rangle$ except that $\delta_g$ is computed
from a foreground population and $\epsilon_i$ from a background population of
galaxies. Because this separation into two populations is usually imperfect,
the overlap allows some contamination from intrinsic alignment. 
In \S \ref{b2bias}, we discuss briefly the inclusion of a nonlinear galaxy-bias.
Finally, we summarize our findings in \S \ref{discuss}. A discussion of the estimators for
the density-ellipticity correlation, in both real and Fourier space, can be found in
the Appendix.

A reading suggestion: for readers interested in the non-Gaussian contributions
to the intrinsic ellipticity-ellipticity and density-ellipticity correlations,
they can skip directly to \S \ref{nonGauss}. Much of \S \ref{setstage} describes
a derivation that, under the assumption of Gaussianity and linear bias, current theories imply a
vanishing intrinsic density-ellipticity correlation -- parts of this are somewhat complicated,
especially those concerning source clustering corrections (\S \ref{intrinsic}), 
and can be skipped by readers not interested in the details; the beginning of
\S \ref{setstage} is still recommended for introducing our notations. 

Before we begin developing our formalism and defining our notations in \S \ref{setstage}, 
it is helpful to mention some previous work on related subjects.
Kaiser (1992) discussed the density-ellipticity correlation in the context of weak-lensing. 
His calculation ignored the contribution from magnification bias. 
Lee \& Pen (2001) discussed galaxy spin-density correlation, but their
focus was on a quantity different from the ones we consider here, and their
emphasis is on application to a three-dimensional survey.
That density-ellipticity correlation might be a useful quantity to
consider in the context of both intrinsic and extrinsic alignments
was mentioned by Catelan et al. (2001) but no explicit calculations
were carried out. Catelan \& Porciani (2001) discussed the density-tidal 
field correlations, but not the density-ellipticity correlation.

\section{Density-Ellipticity Correlation}
\label{setstage}

The ellipticity of a galaxy can be defined using the quadrupole
moment $Q_{ij}$ of the light distribution: $Q_{ij} = 
\int d^2 \theta \theta_i \theta_j I (\bi{\theta})$, where
the origin is chosen to be the centroid of the image, and
$I(\bi{\theta})$ is the intensity profile.
\begin{equation}
\epsilon_1 \equiv {Q_{11} - Q_{22} \over Q_{11} + Q_{22}}
\quad , \quad 
\epsilon_2 \equiv {2 Q_{12} \over Q_{11} + Q_{22}}
\label{epsilonQ}
\end{equation}
This definition of ellipticity is also consistent with
$\epsilon_1 = \epsilon {\,\rm cos\,} 2\alpha$ and
$\epsilon_2 = \epsilon {\,\rm sin\,} 2\alpha$, where
$\epsilon \equiv (a^2 - b^2)/(a^2 + b^2)$, with $a$ and
$b$ being the major and minor axes and $\alpha$ being the
orientation angle of the major axis. 

It is customary to consider the following two different
ellipticity-ellipticity correlations
(or their Fourier transforms, or the corresponding variance). 
Suppose $\Delta\bi{\theta} = 
\Delta\theta $ $({\,\rm cos\,}\psi_\theta , {\,\rm sin\,}\psi_\theta)$
is the angular separation vector between two galaxies (or two pixels
where ellipticity estimates exist), with $\psi_\theta$ being
the orientation angle between the separation vector and
one's chosen x-axis, then one can define the tangential and radial ellipticities:
\begin{eqnarray}
\label{epsilontr}
\epsilon_t = \epsilon_1 {\,\rm cos\,} 2\psi_\theta + \epsilon_2 {\,\rm sin\,} 2\psi_\theta \\ \nonumber
\epsilon_r = -\epsilon_1 {\,\rm sin\,} 2\psi_\theta + \epsilon_2 {\,\rm cos\,} 2\psi_\theta
\end{eqnarray}
which are equivalent to the ellipticities ($\epsilon_1$ and $\epsilon_2$) 
if the x-axis were chosen to lie along
the separation vector. The two corresponding correlations are
$\langle \epsilon_t (\bi{\theta}) \epsilon_t (\bi{\theta'}) \rangle$ and
$\langle \epsilon_r (\bi{\theta}) \epsilon_r (\bi{\theta'}) \rangle$, which
depend only on the separation $\Delta\theta = |\bi{\theta} - \bi{\theta'}|$ 
but not its orientation.
The cross-correlation between tangential and radial ellipticities vanishes
by parity invariance. The electric and magnetic correlation functions
can be constructed from combinations of the above correlations and their
derivatives (see Kamionkowski, \, Kosowsky \& Stebbins 1997, Seljak \& Zaldarriaga 1997). 
	
We are interested, on the other hand, in the density-ellipticity
correlation: $\langle \delta_g (\bi{\theta}) \epsilon_t (\bi{\theta'}) \rangle$,
where $\delta_g$ is the galaxy overdensity. The correlation between
density and the radial ellipticity vanishes by parity. 

The observed ellipticity is divided into intrinsic and extrinsic parts:
\begin{equation}
\epsilon_i = \epsilon_i^{\rm in.} + \epsilon_i^{\rm ex.}
\label{epinex}
\end{equation}
where $i$ can stand for $1$ and $2$, or $t$ and $r$. 

Note that all of the above quantities are observed 
in projection i.e.
\begin{eqnarray}
\label{epsilonproj}
\epsilon_i (\bi{\theta}) 
= \int d\chi_g W_g (\chi_g) \epsilon_i (\bi{\theta}, \chi_g)
\end{eqnarray}
where $\chi_g$ is the comoving distance along the line
of sight, and we use $\epsilon_i$ with an extra argument
of $\chi_g$ to denote the ellipticity at a given angular
position and a particular redshift (and similarly for
$\epsilon_i^{\rm in.}$ and $\epsilon_i^{\rm ex.}$). 
Here $W_g (\chi_g)$ represents the distribution of
galaxies along the line of sight, whose normalization
is $\int d\chi_g W_g (\chi_g) = 1$. We will refer to it
as the selection function.

The galaxy overdensity is similarly projected \footnote{
This expression should in fact be slightly 
modified by magnification bias, which we will discuss
in \S \ref{extrinsic}.}:
\begin{eqnarray}
\label{deltagproj}
\delta_g (\bi{\theta}) 
= \int d\chi_g W_g (\chi_g) \delta_g (\bi{\theta}, \chi_g)
\end{eqnarray}

The metric we adopt is:
\begin{eqnarray}
ds^2 = -dt^2 + a(t)^2 [d\chi^2 + r(\chi)^2 (d\theta^2 + {\,\rm sin}^2 \theta
d\phi^2)]
\end{eqnarray}
where $r(\chi) = K^{-1/2}{\,\rm sin} K^{1/2} \chi$, 
$(-K)^{-1/2} {\,\rm sinh} (-K)^{1/2} \chi$, $\chi$
for a closed, open and flat universe respectively, and
$K = -\Omega_k H_0^2 / c^2$ with $H_0$ being the
Hubble constant today, $c$ is the speed of light and
$\Omega_k$ is the fraction of critical density in curvature; 
$a(t)$ is the Hubble scale factor as a function of proper
time $t$. In this paper, we will display examples exclusively
for the model where the matter density $\Omega_m = 0.3$,
and the vacuum density $\Omega_\Lambda = 0.7$, although
the formalism allows other possibilities (see e.g. Hui 1999,
Benabed \& Bernardeau 2001, Cooray \& Hu 2001, 
Huterer 2001, Hu 2002 for studies of quintessence models in the 
lensing context). 

Before we start describing our calculations
in detail, it is important to emphasize that
because the ellipticity $\epsilon_t$ can only be estimated where there are 
galaxies, the ellipticity is always implicitly weighed by the local galaxy 
density. In other words,
the actual quantity that can be realistically measured is:
\begin{eqnarray}
\label{xicrossIN}
\xi_{\rm cross} (|\bi{\theta} - \bi{\theta'}|) \equiv
\int d\chi_g W_g (\chi_g) 
\int d\chi_g' W_g (\chi_g') \\ \nonumber 
\langle \delta_g (\bi{\theta}, \chi_g)
(1+\delta_g (\bi{\theta'}, \chi_g')) \epsilon_t (\bi{\theta'}, \chi_g') 
\rangle
\end{eqnarray}
For convenience, we define the following quantity:
\begin{eqnarray}
\label{clusteringC}
\epsilon_t^\Delta (\bi{\theta}) \equiv \int d\chi_g W_g(\chi_g) \delta_g (\bi{\theta}, \chi_g) 
\epsilon_t (\bi{\theta}, \chi_g)
\end{eqnarray}
Then, $\xi_{\rm cross} = \langle \delta_g (\bi{\theta}) \epsilon_t (\bi{\theta'}) \rangle
+ \langle \delta_g (\bi{\theta}) \epsilon_t^\Delta (\bi{\theta'}) \rangle$, 
instead of simply $\langle \delta_g (\bi{\theta}) \epsilon_t (\bi{\theta'}) \rangle$.
This sort of correction is sometimes referred to as a source
clustering effect (see Bernardeau 1998 who discussed this effect in the context of extrinsic
ellipticity-ellipticity correlations). 
To be concrete, an estimator for $\xi_{\rm cross}$ is given in the Appendix.
We also provide an estimator for its Fourier analog. 

In this paper, we will ignore the source clustering correction 
in the context of
lensing (as is commonly done for 
ellipticity-ellipticity correlations). 
On the other hand, 
as far as intrinsic alignment is concerned, since, as we will see,
the lowest order term $\langle \delta_g \epsilon_t \rangle$ vanishes (according
to Gaussian theories),
we will consider the term $\langle \delta_g \epsilon_t^\Delta \rangle$ as well. 
We will use the symbol ${\epsilon_t^{\Delta\,\rm in.}}$ to denote
the intrinsic part of $\epsilon_t^\Delta$. 

\subsection{Intrinsic Alignment -- Gaussian Theories}  
\label{intrinsic}

We calculate here both $\langle \delta_g \epsilon_t^{\rm in.} \rangle$
and $\langle \delta_g {\epsilon_t^{\Delta\,\rm in.}} \rangle$ using
two different formulations of tidal alignment theories.

\subsubsection{Tidal Alignment Theory According to MWK}
\label{theoryI}

We start with the simpler theory, developed by MWK
(see also Catelan et al. 2001), who postulated that
\begin{eqnarray}
\label{mackey}
&& \epsilon_1^{\rm in.} 
= {\beta \over 15} [(2\phi_{33} - \phi_{11} - \phi_{22})(\phi_{11}-\phi_{22})
\\ \nonumber 
&& + 3(\phi_{23}^2 - \phi_{13}^2)] \\ \nonumber
&& \epsilon_2^{\rm in.} 
= {\beta \over 15} [2(2\phi_{33}-\phi_{11}-\phi_{22})\phi_{12}
- 6 \phi_{13} \phi_{23}]
\end{eqnarray}
where $\beta$ is a constant which quantifies the mean ellipticity 
of galaxies. The above is motivated by the tidal torquing theory for the origin
of angular momentum developed by Peebles (1969), Doroshkevich (1970) and White (1984), and 
assuming the moment of inertia tensor is uncorrelated with the local
tidal field. 
Here, the third direction is taken to be along the line of sight,
$\phi$ is the gravitational potential at some given point
in space, and $\phi_{ij} \equiv \nabla_i \nabla_j \phi$. 

We relate $\delta_g$ to $\phi$ using the linear bias model
and the Poisson equation i.e.
\begin{eqnarray}
\label{linearbias}
&& \delta_g = b \delta \\ \nonumber
&& \nabla^2 \phi = 4\pi G \bar\rho a^2 \delta = (3H_0^2 \Omega_m /2/a) \delta
\end{eqnarray}
where $\delta$ is the mass overdensity ($\delta = (\rho - \bar \rho)/\bar \rho$
with $\rho$ being the mass density and $\bar\rho$ its mean), and
$b$ is the bias factor, which can be redshift dependent in general
(Kaiser 1984, Bardeen et al. 1986, Fry \& Gaztanaga 1993, Mo \& White 1996).
\footnote{Recent more sophisticated modeling of galaxy biasing using the halo model 
still respects linearity on large scales (see e.g. Peacock \& Smith 2000, Seljak 2000,
Ma \& Fry 2000, Scoccimarro et al. 2001, Berlind \& Weinberg 2001).
See also Tegmark et al. (2001) for a measurement of
the galaxy-mass correlation coefficient, which is consistent
with a deterministic linear bias on large scales. 
Note that the tidal theories proposed by MWK and 
CNPT are also expected to be valid only
on large scales.}

Limber's approximation, which assumes that the selection function
is slowly varying compared to the correlation function (Peebles 1980), 
tells us how to relate
projected correlations to their three-dimensional counterparts:
\begin{eqnarray}
\label{Limber}
&& \langle \delta_g (\bi{\theta}) \epsilon_i (\bi{\theta'}) \rangle
= \int d\chi_g W_g^2 \int dp_3 \langle \delta_g ({\bi p}) \epsilon_i ({\bi q}) \rangle
\\ \nonumber 
&& \langle \delta_g (\bi{\theta}) \epsilon_i^\Delta (\bi{\theta'}) \rangle
= \int d\chi_g W_g^2 \int dp_3 \langle \delta_g ({\bi p}) \epsilon_i ({\bi q}) \delta_g ({\bi q})\rangle
\end{eqnarray}
where the selection function $W_g(\chi_g)$ is as defined in eq. (\ref{epsilonproj}) \&
(\ref{deltagproj}), 
${\bf p}$ and ${\bf q}$ are three-dimensional positions, 
the z-axis (labeled $3$) is the direction along the line of sight,
and ${\bf p_\perp} - {\bf q_\perp}$ (components of
${\bf p}$ and ${\bf q}$ perpendicular to $z$) is chosen such that
${\bf p_\perp} - {\bf q_\perp} = {\bi{ \Delta \theta}} r(\chi_g)$. 

From eq. (\ref{linearbias}) and (\ref{mackey}), it
is easy to see that
\begin{eqnarray}
\label{deltaepsilonAB}
\langle \delta({\bf p}) \epsilon_i^{\rm in} ({\bf q}) \rangle
= \sum_{A, B} c_{AB}
\langle \nabla^2 \phi ({\bf p}) \phi_A ({\bf q}) \phi_B ({\bf q}) \rangle
\end{eqnarray}
where $A$ and $B$ signify various second derivatives, and
$c_{AB}$ represents coefficients independent of spatial positions
(except through the overall redshift). 
That $\phi$ has equal probability of being positive or negative
implies the above must vanish (for instance, for a Gaussian random
distribution of $\phi$, as is assumed by MWK; see \S \ref{nonGauss} for a discussion
of expected violations of this assumption). 
That the above vanishes for both $i = 1$ and $i = 2$ implies
$\langle \delta_g (\bi{\theta}) \epsilon_t (\bi{\theta'}) \rangle$ must be zero 
as well when a projection is applied. 

How about 
$\langle \delta_g ({\bf p}) \epsilon_i^{\rm in.} ({\bf q}) \delta_g ({\bf q}) \rangle$, 
which shows up in the source clustering correction (eq. [\ref{Limber}])? 
This requires a little bit more work. From eq. (\ref{mackey}), it is
clear there are several terms. We will work out the case of 
$i=2$ in detail. The case of $i=1$ is very similar, albeit with more terms.
We have $\langle \delta_g \epsilon_2^{\rm in.} \delta_g \rangle 
= b^2 \langle \delta \epsilon_2^{\rm in.} \delta \rangle$, and the latter factor
is given by:

\begin{eqnarray}
\label{deltaepdelta}
\langle \delta ({\bf p}) \epsilon_2^{\rm in.} ({\bf q}) \delta ({\bf q}) \rangle
= {2 \beta \over 15} [ 2 \langle \delta ({\bf p}) \delta ({\bf q}) \phi_{33} ({\bf q}) 
\phi_{12} ({\bf q}) \rangle \\ \nonumber
- \langle \delta ({\bf p}) \delta ({\bf q}) \phi_{11} ({\bf q}) \phi_{12} ({\bf q}) \rangle \\ \nonumber 
- \langle \delta ({\bf p}) \delta ({\bf q}) \phi_{22} ({\bf q}) \phi_{12} ({\bf q}) \rangle \\ \nonumber
- 3 \langle \delta ({\bf p}) \delta ({\bf q}) \phi_{13} ({\bf q}) \phi_{23} ({\bf q}) \rangle ]
\end{eqnarray}

By MWK's assumption of a Gaussian random field $\phi$, each of
the above terms factorizes into products of second moments. 
For instance, the first term gives
\begin{eqnarray}
\label{firsterm}
2 \langle \delta ({\bf p}) \delta ({\bf q}) \phi_{33} ({\bf q}) 
\phi_{12} ({\bf q}) \rangle
= 2 \langle \delta ({\bf p}) \phi_{33} ({\bf q}) \rangle \langle \delta ({\bf q}) \phi_{12} ({\bf q}) \rangle
\\ \nonumber 
+ 2 \langle \delta ({\bf p}) \phi_{12} ({\bf q}) \rangle \langle \delta ({\bf q}) \phi_{33} ({\bf q}) \rangle
\\ \nonumber 
+ 2 \langle \delta ({\bf p}) \delta ({\bf q}) \rangle \langle \phi_{12} ({\bf q}) \phi_{33} ({\bf q}) \rangle
\end{eqnarray}

Several of the terms above contain an average of the following form:
\begin{eqnarray}
\langle \delta ({\bf p}) \phi_{ij} ({\bf q}) \rangle 
= {1 \over 4\pi G \bar \rho a^2} \int {d^3 k \over (2\pi)^3} P_\phi (k) k^2 k_i k_j e^{-i{\bf k} \cdot ({\bf p} - {\bf q})}
\label{deltaphi}
\end{eqnarray}
where $P_\phi$ is the potential power spectrum in the sense that
$\langle \phi({\bf p}) \phi ({\bf q}) \rangle = \int (d^3 k / [2\pi]^3) P_\phi (k) {\,\rm exp\,}
[-i {\bf k} \cdot ({\bf p} - {\bf q})]$. 

The above expression 
can also be used to evaluate $\langle \delta ({\bf q}) \phi_{ij} ({\bf q}) \rangle$
by setting ${\bf p} = {\bf q}$. In fact, doing so, it is clear that
$\langle \delta ({\bf q}) \phi_{12} ({\bf q}) \rangle$ vanishes
by isotropy (i.e. $P_\phi (k)$ depends on magnitude of $k$ but not its direction). 
We can therefore ignore the first term on the right hand side of
eq. (\ref{firsterm}). 

Similarly, a term like $\langle \phi_{33} ({\bf q}) \phi_{12} ({\bf q}) \rangle$ must
vanish, because this is given by
\begin{eqnarray}
\langle \phi_{33} ({\bf q}) \phi_{12} ({\bf q}) \rangle 
= \int {d^3 k \over (2\pi)^3} P_\phi (k) k_3^2 k_1 k_2 
\end{eqnarray}
which vanishes by isotropy.

Therefore, the only term that would eventually survive from eq. (\ref{firsterm}) 
is $2 \langle \delta ({\bf p}) \phi_{12} ({\bf q}) \rangle \langle \delta ({\bf q}) \phi_{33} ({\bf q}) \rangle$. This is (ignoring the factor of
$2\beta /15$) from the first term on the right side of eq. (\ref{deltaepdelta}).
Most of the rest of the terms from eq. (\ref{deltaepdelta}) vanish due to the same reasons as above
except for two terms (which come from the second and third terms
in eq. [\ref{deltaepdelta}]):
$-\langle \delta ({\bf p}) \phi_{12} ({\bf q}) \rangle
\langle \delta ({\bf q}) \phi_{11} ({\bf q}) \rangle 
- \langle \delta ({\bf p}) \phi_{12} ({\bf q}) \rangle
\langle \delta ({\bf q}) \phi_{22} ({\bf q}) \rangle$. 
Finally, collecting all the surviving terms, it is easy to see that
\begin{eqnarray}
&& 2 \langle \delta ({\bf p}) \phi_{12} ({\bf q}) \rangle 
\langle \delta ({\bf q}) \phi_{33} ({\bf q}) \rangle
\\ \nonumber 
&& -\langle \delta ({\bf p}) \phi_{12} ({\bf q}) \rangle
\langle \delta ({\bf q}) \phi_{11} ({\bf q}) \rangle \\ \nonumber 
&& - \langle \delta ({\bf p}) \phi_{12} ({\bf q}) \rangle
\langle \delta ({\bf q}) \phi_{22} ({\bf q}) \rangle
= 0
\end{eqnarray}
because $\langle \delta ({\bf q}) \phi_{33} ({\bf q}) \rangle
= \langle \delta ({\bf q}) \phi_{22} ({\bf q}) \rangle = $
$\langle \delta ({\bf q}) \phi_{11} ({\bf q}) \rangle$.
This follows again from isotropy.

Therefore, the conclusion is that 
$\langle \delta ({\bf p}) \epsilon_2^{\rm in.} ({\bf q}) \delta ({\bf q}) \rangle$
as given in eq. (\ref{deltaepdelta}) vanishes exactly. The same statement can be proven for
$\langle \delta ({\bf p}) \epsilon_1^{\rm in.} ({\bf q}) \delta ({\bf q}) \rangle$,
using very similar arguments as those outlined above.
Hence, in summary, we find that (using the above together
with eq. [\ref{Limber}])
{\it according to the tidal theory of MWK, and under
the assumption of a linear bias, both 
$\langle \delta_g (\bi{\theta}) \epsilon_i (\bi{\theta'}) \rangle$
and its source clustering correction 
$\langle \delta_g (\bi{\theta}) \epsilon_i^\Delta (\bi{\theta'}) 
\rangle$ vanish exactly for both $i=1$ and $i=2$, and therefore, for $i= t$ as well.}

\subsubsection{Tidal Alignment Theory According to CNPT}
\label{theory2}

Next, we turn our attention to the tidal theory of CNPT
(which built upon earlier work by Lee \& Pen 2000), according to which
the galaxy ellipticity is given by:
\begin{eqnarray}
\label{crittenden}
&& \epsilon_1^{\rm in.} = {\alpha \over 2} \sum_{i=1}^3 
(\hat \phi_{1i} \hat \phi_{i1} - \hat \phi_{2i} \hat \phi_{i2}) \\ \nonumber
&& \epsilon_2^{\rm in.} = \alpha 
\sum_{i=1}^3\hat \phi_{1i} \hat \phi_{i2}
\end{eqnarray}
where $\alpha$ is a constant which depends on both the average ellipticity
of galaxies and the degree to which the moment of inertia tensor
is correlated with the local tidal field
\footnote{The symbol $\alpha$ here is actually equal to $a\alpha (6 - 9\pi/4)$ in 
the notation of CNPT.}, $\hat \phi_{ij}$
is the unit normalized traceless tidal tensor i.e.
$\hat \phi_{ij} \equiv 
[\nabla_i \nabla_j \phi  - {1\over 3} \delta_{ij} \nabla^2 \phi] / N$, 
where $N$ is a normalizing factor so that 
$\sum_{i,j} \hat \phi_{ij} \hat \phi_{ij} = 1$
(summation of $i,j$ is over $1,2,3$). As such, this theory
relates the galaxy ellipticity to the direction of the galaxy angular
momentum, but not its amplitude. 
\footnote{Porciani et al. (2001a,b) recently showed using numerical 
simulations that much of the angular momentum
correlation in dark matter halos induced by tides is erased by non-linear effects.
One can think of modeling this by decreasing $\alpha$ in eq. (\ref{crittenden}).
See also van den Bosch et al. (2002).}

Here, we would like to evaluate the density-ellipticity correlation
using Limber's approximation and linear bias just as we have done
using the other tidal theory (eq. [\ref{Limber}] \& [\ref{linearbias}]).
We therefore, need to compute $\langle \delta ({\bf p}) \epsilon_i^{\rm in.} ({\bf q}) \rangle$
and $\langle \delta ({\bf p}) \epsilon_i^{\rm in.} ({\bf q}) \delta ({\bf q})\rangle$
at two spatial points ${\bf p}$ and ${\bf q}$ just as before.
To do so, we need the two-point
probability distribution for $\phi_{ij}$. Adopting the notation
$T \equiv (\phi_{11}, \phi_{22}, \phi_{33}, \phi_{12}, \phi_{13}, \phi_{23})$
where $T$ signifies the whole vector of different components of
the tidal field (i.e. $T_1 = \phi_{11}$, etc.), 
the two-point probability distribution used by CNPT is
Gaussian random i.e. 
\begin{eqnarray}
\label{gauss}
P(T ({\bf p}), T ({\bf q})) = 
{1 \over \sqrt{{\,\rm det\,} C}
(2\pi)^6 }  {\,\rm exp\,}\left[
- {1\over 2} {\vec T}^T C^{-1} {\vec T} \right] \nonumber
\end{eqnarray}
where $P(T ({\bf p}), T ({\bf q})) d^6 T ({\bf p}) d^6 T ({\bf q})$
gives the probability that at spatial points 
${\bf p}$ and ${\bf q}$, the tidal field vectors
take their respective values in the above ranges.
Here $\vec T \equiv (T ({\bf p}), T ({\bf q}))$, and
${\vec T}^T$ is its transpose. 
The matrix $C$ gives the correlation matrix, which has the block 
diagonal form:
\begin{eqnarray}
C = \left[ \begin{array}{cc} C_0 & C_{\bf p - q} \\
C_{\bf p - q} & C_0
\end{array} \right]
\end{eqnarray}
where $C_0$ is the $6 \times 6$ zero-lag correlation matrix, and
$C_{\bf p - q}$ is the $6 \times 6$ two-point correlation matrix. 

Let us first evaluate $\langle \delta \epsilon_i^{\rm in.} \rangle$. 
\begin{eqnarray}
\label{dep00}
\langle \delta ({\bf p}) \epsilon_i^{\rm in.} ({\bf q}) \rangle
= \sum_{A,B} d_{AB} \langle \nabla^2 \phi ({\bf p}) 
\hat \phi_A ({\bf q}) 
\hat \phi_B ({\bf q}) \rangle
\end{eqnarray}
where $A$ and $B$ denote appropriate double indices as given
in eq. (\ref{crittenden}), 
and $d_{AB}$ represents coefficients independent of spatial positions
(except through the overall redshift). 

Clearly, the probability
distribution in eq. (\ref{gauss}) is invariant
under $T \rightarrow -T$ (i.e. $\phi \rightarrow -\phi$), while
the combination $\nabla^2 \phi \hat \phi_A \hat \phi_B$ (eq. 
[\ref{dep00}]) switches
sign under such a transformation. Therefore, the expectation value 
$\langle \delta ({\bf p}) \epsilon_i^{\rm in.} ({\bf q}) \rangle$
must vanish. 

Next, let us consider:
\begin{eqnarray}
\label{ded}
&& \langle \delta ({\bf p}) \epsilon_i^{\rm in.} ({\bf q}) \delta ({\bf q}) \rangle
\\ \nonumber 
&& = \sum_{A,B} e_{AB} 
\langle \nabla^2 \phi ({\bf p}) \hat \phi_A ({\bf q}) \hat \phi_B ({\bf q})
\nabla^2 \phi ({\bf q}) \rangle \\ \nonumber
&& = \sum_{A,B} e_{AB} \langle {\,\rm Tr\,} T ({\bf p}) {\,\rm Tr\,} T({\bf q})
\hat T_A ({\bf q}) \hat T_B ({\bf q}) \rangle
\end{eqnarray}
where $\hat T \equiv (\hat \phi_{11}, \hat \phi_{22}, \hat \phi_{33},
\hat \phi_{12}, \hat \phi_{13}, \hat \phi_{23})$, 
${\,\rm Tr\,} T \equiv T_1 + T_2 + T_3$, and 
$e_{AB}$ represents some coefficient. 
Following CNPT, 
it is useful to rotate 
the vector $T$ by defining ${\cal T} \equiv R T$, where
$R$ is an invertible matrix such that ${\cal T} = 
({\rm Tr\,}T/\sqrt{3}, (T_{1}-T_{2})/\sqrt{2}, (T_{1}+T_{2} - 2 T_{3})/\sqrt{6}, 
\sqrt{2} T_{4}, \sqrt{2} T_{5}, $ $\sqrt{2} T_{6})$.
With this rotation, the above expression can be rewritten as
\begin{eqnarray}
\label{eprimeAB}
&& \langle \delta ({\bf p}) \epsilon_i^{\rm in.} ({\bf q}) \delta ({\bf q}) \rangle
\\ \nonumber 
&& = \sum_{A,B} e'_{AB} \langle {\cal T}_1 ({\bf p}) {\cal T}_1 ({\bf q})
\hat {\cal T}_A ({\bf q}) \hat {\cal T}_B ({\bf q}) \rangle
\end{eqnarray}
where $e'_{AB} = \sum_{D,E} e_{DE} R^{-1}_{DA} R^{-1}_{EB}$, and
$\hat {\cal T} \equiv R \hat T$. 
Note that because $\hat T$ is formed from the traceless part of the
tidal tensor, $\hat {\cal T}_1 = 0$ be definition. 
It is also useful to define $|{\cal T}| \equiv 
[{\cal T}_2^2 + {\cal T}_3^2 + ... + {\cal T}_6^2]^{1/2}$, so
that ${\cal T}_A = \hat {\cal T}_A |{\cal T}|$ for $A \ne 1$. 

The advantage of such a rotation is that the 
correlation matrix at zero-lag becomes diagonalized:
${\cal C}_0 \equiv R C_0 R^T$ is diagonal: 
${\cal C}_0 = (\xi_0/15) {\,\rm diag\,} (5,2,2,2,2,2)$, where
$\xi_0$ is the zero-lag correlation of ${\,\rm Tr\,}T$. 
Similarly, we define ${\cal C}_{\bf p - q} \equiv
R C_{\bf p - q} R^T$, and 
\begin{eqnarray}
{\cal C} \equiv \left[ \begin{array}{cc} {\cal C}_0 & {\cal C}_{\bf p - q} \\
{\cal C}_{\bf p - q} & {\cal C}_0
\end{array} \right]
\end{eqnarray}
but note that neither ${\cal C}_{\bf p - q}$ nor ${\cal C}$ is diagonal
in general. 

The expectation value $\langle {\cal T}_1 {\cal T}_1 \hat {\cal T}_A
\hat {\cal T}_B \rangle $ can be written as follows. 
\begin{eqnarray}
\label{stuff}
&&\langle {\cal T}_1 ({\bf p}) {\cal T}_1 ({\bf q}) \hat {\cal T}_A ({\bf q})
\hat {\cal T}_B ({\bf q}) \rangle \\ \nonumber 
&&= \int 
{d^6 {\cal T} ({\bf p}) d^6 {\cal T} ({\bf q})
\over \sqrt{{\,\rm det\,} {\cal C}} (2\pi)^6}
{\cal T}_1 ({\bf p}) {\cal T}_1 ({\bf q}) \hat {\cal T}_A ({\bf q})
\hat {\cal T}_B ({\bf q}) 
\\ \nonumber && \quad \times {\,\rm exp}
\left[ - {1\over 2} \vec {\cal T}^T {\cal C}^{-1} {\cal T} \right]
\\ \nonumber 
&&= - \int 
{d^6 {\cal T} ({\bf p}) d^6 {\cal T} ({\bf q})
\over \sqrt{{\,\rm det\,} {\cal C}} (2\pi)^6}
\hat {\cal T}_A ({\bf q})
\hat {\cal T}_B ({\bf q}) 
\\ \nonumber && \quad \times 
{\partial \over \partial {\cal C}^{-1}_{17}} {\,\rm exp}
\left[ - {1\over 2} \vec {\cal T}^T {\cal C}^{-1} {\cal T} \right]
\\ \nonumber 
&& = - {\partial \over \partial {\cal C}^{-1}_{17}} \langle 
\hat {\cal T}_A ({\bf q})
\hat {\cal T}_B ({\bf q}) \rangle \\ \nonumber
&& \quad + {\partial ({\,\rm det\,} {\cal C})^{-1/2}
\over \partial {\cal C}^{-1}_{17}} \sqrt{{\,\rm det\,} {\cal C}}
\langle 
\hat {\cal T}_A ({\bf q})
\hat {\cal T}_B ({\bf q}) \rangle 
\end{eqnarray}
where ${\cal C}^{-1}_{17}$ represents the $(1,7)$ component
of the symmetric matrix ${\cal C}$. 
Note that $d^6 {\cal T} = d{\,\rm Tr\,}T |{\cal T}|^4 d |{\cal T}| d^4 \hat {\cal T}
/\sqrt{3}$. 

The second term in the last equality above
(the one involving a derivative of the determinant), when contracted with $e'_{AB}$ in 
eq. (\ref{eprimeAB}), gives us something that is proportional
to $\langle \epsilon_i^{\rm in.} ({\bf q}) \rangle$, which
must vanish by isotropy. Therefore, all we have to
worry about is the first term. At first sight, it appears
that the same argument might apply: after contracting
it with $e'_{AB}$, the first term gives a derivative
of $\langle \epsilon_i^{\rm in.} \rangle$ with respect to ${\cal C}^{-1}_{17}$.
Should this also vanish by isotropy? 
The ultimate answer will turn out to be yes, but
one has to reason with some care: does the fact that
we are varying ${\cal C}^{-1}_{17}$ while keeping the
rest of the components of ${\cal C}$ fixed somehow
spoil isotropy? One would expect not, because the component
we are varying has to do with the trace of the tidal tensor. 
However, let us prove this explicitly. 

The matrix ${\cal C}^{-1}$ in general takes the following
form:
\begin{eqnarray}
{\cal C}^{-1} = \left[ \begin{array}{cc} -{\cal C}^{-1}_{\bf p-q}
{\cal C}_0 M & M \\
M & - {\cal C}^{-1}_{\bf p-q} {\cal C}_0 M 
\end{array} \right]
\end{eqnarray}
where $M$ is some matrix that satisfies
\begin{eqnarray}
\label{M}
- {\cal C}_0 {\cal C}^{-1}_{\bf p-q} {\cal C}_0 M + {\cal C}_{\bf p-q}
M = 1
\end{eqnarray}
to ensure ${\cal C} {\cal C}^{-1}$ gives the identity. 

When we are varying ${\cal C}^{-1}_{17}$, we are therefore varying
$M_{11}$ while keeping everything else fixed. 
This means that $\Delta ({\cal C}^{-1}_{\rm p-q} {\cal C}_0 M) = 0$. 
Eq. (\ref{M}) should remain valid under the variation, and so
$-\Delta ({\cal C}_0 
{\cal C}^{-1}_{\rm p-q} {\cal C}_0 M) + \Delta ({\cal C}_{\bf p -q}
M) = 0$. Combining both, together with eq. (\ref{M}), it can be shown that varying
${\cal C}^{-1}_{17}$ implies a change in ${\cal C}_0$ of the form
\begin{eqnarray}
\Delta {\cal C}_0 = - {\cal C}_{\bf p-q} \Delta M 
{\cal C}_0 - {\cal C}_0 \Delta M {\cal C}_{\bf p-q}
\end{eqnarray}
The fact that the only non-zero component of $\Delta M$ is $\Delta M_{11}
= \Delta {\cal C}^{-1}_{17}$,
together with the fact that ${\cal C}_0$ is diagonal, implies
that $[\Delta {\cal C}_0 ]_{ij} = 0$ except for $i$ or $j$ (or both) $=1$. 
This is an important fact we will make use of later.

Dropping the ${\bf q}$ label for simplicity, let us then consider
the term from eq. (\ref{stuff}):
\begin{eqnarray}
\label{partialC}
&& - {\partial \over \partial {\cal C}^{-1}_{17}}
\langle \hat {\cal T}_A \hat {\cal T}_B \rangle
\\ \nonumber 
&& = - {\partial \over \partial {\cal C}^{-1}_{17}}
\int {d^6 {\cal T} \over \sqrt{{\,\rm det\,} {\cal C}_0}
(2\pi)^3} \hat {\cal T}_A \hat {\cal T}_B
{\,\rm exp\,} \left[ - {1\over 2} {\cal T}^T {\cal C}_0^{-1} {\cal T} \right]
\\ \nonumber 
&& = Q - (\Delta {\cal C}_{17}^{-1})^{-1}  \int {d^6 {\cal T} \over \sqrt{{\,\rm det\,} {\cal C}_0}
(2\pi)^3} \hat {\cal T}_A \hat {\cal T}_B \\ \nonumber 
&& \times \left[{1\over 2} {\cal T}^T {\cal C}_0^{-1} \Delta {\cal C}_0 {\cal C}_0^{-1}
{\cal T} \right]
{\,\rm exp\,} \left[ - {1\over 2} {\cal T}^T {\cal C}_0^{-1} {\cal T} \right]
\end{eqnarray}
where $Q$ is a term that involves the derivative of the determinant
${\,\rm det\,} {\cal C}_0$ with respect to ${\cal C}^{-1}_{17}$, and
$Q$ is proportional to $\langle \hat {\cal T}_A \hat {\cal T}_B \rangle$,
which as we have argued before, vanishes by isotropy under contraction with
$e'_{AB}$ (eq. [\ref{eprimeAB}]) because it gives something
that is proportional to $\langle \epsilon_i^{\rm in.} \rangle$. 

Now, recall the fact that $(\Delta {\cal C}_0)_{ij}$
is non-zero only if $i$ or $j$ $=1$. This, together with the fact
that ${\cal C}_0^{-1}$ is diagonal, implies that aside from $Q$ (which
we can ignore), all terms in eq. (\ref{partialC}) are of the form
$\eta \langle \hat {\cal T}_A \hat {\cal T}_B ({\cal T}_1)^2 \rangle$
or $\sum_C \zeta_C \langle \hat {\cal T}_A \hat {\cal T}_B {\cal T}_1 \hat {\cal T}_C |{\cal T}|\rangle$
where $\eta$ and $\zeta_C$ are some coefficients.
The fact that the latter term involves an odd number of directional
vectors $\hat {\cal T}$ implies it must be zero. 

Therefore, the only term we need to consider is
\begin{eqnarray}
\langle \delta ({\bf p}) \epsilon_i^{\rm in.} ({\bf q}) \delta ({\bf q}) \rangle 
= \sum_{A,B} \eta e'_{AB} \langle \hat {\cal T}_A ({\bf q}) \hat {\cal T}_B ({\bf q}) 
[{\,\rm Tr \,} T ({\bf q})]^2 \rangle \\ \nonumber
= \sum_{A,B} \eta e_{AB} \langle \hat T_A ({\bf q}) \hat T_B ({\bf q}) 
[{\,\rm Tr \,} T ({\bf q})]^2 \rangle 
\end{eqnarray}
Note that $\eta$ is in principle dependent on ${\bf p -q}$. 
The coefficients $e_{AB}$ are determined by the expressions for
$\epsilon_i^{\rm in.}$ in eq. (\ref{crittenden}). 
For $i=1$, the above is proportional to 
$\langle (\hat T_{1} \hat T_1 - \hat T_{2} \hat T_2 ) ({\,\rm Tr\,} T)^2 \rangle
+ \langle (\hat T_{5} \hat T_5 - \hat T_6 \hat T_6 ) ({\,\rm Tr\,} T)^2 \rangle$.
The important point to keep in mind is that this expectation value
is evaluated at a single point ({\bf q}), and with no preferred direction,
this clearly vanishes by isotropy. 
Similarly, the $i=2$ term vanishes as well. 

To summarize, we find that {\it according to the tidal theory formulated
by CNPT, and under the assumption of linear bias, both
$\langle \delta_g (\bi{\theta}) \epsilon_i (\bi{\theta'}) \rangle$ and 
its source clustering correction
$\langle \delta_g (\bi{\theta}) \epsilon_i^\Delta (\bi{\theta'}) \rangle$ 
vanish exactly.} This holds for $i = 1$, $i=2$ and therefore for $i=t$ as well.

\subsection{Extrinsic Alignment -- Gravitational Lensing}
\label{extrinsic}

Gravitational lensing induces a correlation between
ellipticity and galaxy density, which we will work out 
in this section. The projected ellipticity and galaxy density
are given in eq. (\ref{epsilonproj}) and eq. (\ref{deltagproj}).
The latter equation needs to be slightly modified to take
into account magnification bias (e.g. Broadhurst, Taylor \& Peacock 1995; Moessner, 
Jain \& Villumsen 1998):
\begin{eqnarray}
\label{deltagproj2}
\delta_g (\bi{\theta})
= \int d\chi_g W_g (\chi_g) [\delta_g (\bi{\theta}, \chi_g) + 5 (s-0.4) \kappa(\bi{\theta}, \chi_g)]
\end{eqnarray}
where $s$ is the slope of the luminosity function in the following sense:
if $N(m,z)$ gives the surface number density of galaxies per unit
magnitude $m$ per unit redshift $z$, $d{\,\rm log\,}N(m,z)/dm = s$. 
Note that $s$ can be redshift dependent, and is the slope
at the faint end of a flux-limited survey. 
Here, $\kappa$ is the lensing convergence, it is related to 
the lensing magnification $|A|$ by $|A| = 1 + 2\kappa$ in
the weak lensing limit. It can be viewed as a projected overdensity
(e.g. Kaiser \& Squires 1993):
\begin{eqnarray}
\label{kappa}
\kappa (\bi{\theta}, \chi_g) = \int_0^{\chi_g} 
{d\chi_L \over a} W_L (\chi_L, \chi_g) \delta (\bi{\theta}, \chi_L)
\end{eqnarray}
where $W_L (\chi_L, \chi_g) = 
(3/2c^2) H_0^2 \Omega_m r(\chi_L) r(\chi_g - \chi_L) / r(\chi_g)$
is the lensing efficiency function. 

Eq. (\ref{deltagproj2}) tells us how the observed number density of
galaxies is modulated by gravitational lensing: magnification allows
more galaxies to be observed in a flux-limited sample, but also causes
the galaxies to appear more spread out. Which effect wins depends on
the slope of the luminosity function. 

Gravitational lensing, of course, also modifies 
the observed galaxy ellipticity: $\epsilon_i = \epsilon_i^{\rm in.}
+ \epsilon_i^{\rm ex.}$ (eq. [\ref{epinex}]). 
We have shown in \S \ref{intrinsic} that $\langle \delta_g \epsilon_i^{\rm in.} \rangle$
vanishes according to two different tidal theories of intrinsic
alignment. Here, we will work out $\langle \delta_g \epsilon_i^{\rm ex.} \rangle$,
for $i = t$ (correlation vanishes for $i=r$, by symmetry under parity). 
\footnote{Since, as we will see, $\langle \delta_g \epsilon_i^{\rm ex.} \rangle$
is non-zero in general, we will not consider in this paper the smaller source
clustering correction: $\langle \delta_g \epsilon_i^{\Delta\, \rm ex.} \rangle$
(which is also often ignored in the case of the ellipticity-ellipticity correlation).}
\footnote{Strictly speaking, one should consider an 
additional cross-correlation between the magnification bias
term involving $\kappa$ and the intrinsic ellipticity $\epsilon_i^{\rm in.}$. 
Since $\kappa$ is simply a projected overdensity $\delta$, the same arguments in 
\S \ref{intrinsic}
apply directly to this cross-correlation as well.}
The lensing induced ellipticity for a galaxy at 
angular position $\bi{\theta}$ and distance $\chi_g$ away 
is given by
\begin{eqnarray}
\label{epsilonex}
\epsilon_i^{\rm ex.} (\bi{\theta}, \chi_g) = D^\theta_i \int_0^{\chi_g}
{d\chi_L \over a} W_L (\chi_L, \chi_g) \tilde \psi (\bi{\theta}, \chi_L)
\end{eqnarray}
where $D^\theta_1 \equiv \nabla^2_{\theta^1} - \nabla^2_{\theta^2}$, 
$D^\theta_2 \equiv 2 \nabla_{\theta^1} \nabla_{\theta^2}$, and
$\tilde \psi (\bi{\theta}, \chi_L) \equiv 2 \phi (\bi{\theta}, \chi_L)/
r(\chi_L)^2 / (4\pi G \bar \rho a^2)$ ($\phi$ is the gravitational
potential that satisfies the Poisson equation as given in eq. [\ref{linearbias}]). 

It is important to note that our definition of $\epsilon^{\rm ex.}_i$ is higher
than what is usually known as the shear, $\gamma_i$, by a factor of $2$.
(The usual convention is that if $\kappa = \nabla_\theta^2 \psi /2$ where
$\psi$ is some projected potential, then $\gamma_1 = (\nabla_{\theta_1}^2 - \nabla_{\theta_2}^2 )
\psi /2$, and $\gamma_2 = \nabla_{\theta_1} \nabla_{\theta_2} \psi$.) 
The reason for our choice is that for simple estimators for the quadrupole
moment $Q_{ij}$, the ellipticity as defined in eq. (\ref{epsilonQ}) is influenced
by lensing according to $\epsilon_i = \epsilon_i^{\rm in.} + \epsilon_i^{\rm ex.} 
= \epsilon_i^{\rm in.} + 2 \gamma_i$, at least
to the lowest order in ellipticity and shear (see Kaiser \& Squires 1993). 

Combining eq. (\ref{epsilonproj}), (\ref{deltagproj2}), (\ref{kappa}) and (\ref{epsilonex}),
together with the definition of tangential ellipticity in eq. (\ref{epsilontr}), 
we obtain
\begin{eqnarray}
\label{exdeltagepsilon}
\langle \delta_g (\bi{\theta}) \epsilon_t^{\rm ex.} (\bi{\theta'}) \rangle
= \langle \delta_g (\bi{\theta}) \epsilon_t^{\rm ex.} (\bi{\theta'}) \rangle_A + 
\langle \delta_g (\bi{\theta}) \epsilon_t^{\rm ex.} (\bi{\theta'}) \rangle_B \\ \nonumber
\langle \delta_g (\bi{\theta}) \epsilon_t^{\rm ex.} (\bi{\theta'}) \rangle_A
\equiv 2 \int {d\chi_L \over a^2} 
{\tilde W_L (\chi_L) \tilde W_L^s (\chi_L) \over r(\chi_L)^2} 
\\ \nonumber 
\int_0^\infty {\ell d\ell \over 2\pi} P(k = \ell / r(\chi_L)) J_2 (\ell |\bi{\theta}
- \bi{\theta'} |) \\ \nonumber 
\langle \delta_g (\bi{\theta}) \epsilon_t^{\rm ex.} (\bi{\theta'}) \rangle_B
\equiv 2 \int {d\chi_L \over a} {\tilde W_L^b (\chi_L) W_g (\chi_g = \chi_L) 
\over r(\chi_L)^2} \\ \nonumber 
\int {\ell d\ell \over 2 \pi} P(k = \ell/r(\chi_L)) J_2 (\ell
|\bi{\theta} - \bi{\theta'} |)
\end{eqnarray}
where 
\begin{eqnarray}
\label{WLs}
\tilde W_L (\chi_L) \equiv \int_{{\chi}_L}^\infty d\chi_g W_g (\chi_g) W_L(\chi_L, \chi_g) \\ \nonumber
\tilde W_L^s (\chi_L) \equiv \int_{\chi_L}^\infty d\chi_g 5(s-0.4) 
W_g (\chi_g) W_L(\chi_L, \chi_g) \\ \nonumber
\tilde W_L^b (\chi_L) \equiv b \int_{\chi_L}^\infty
d\chi_g W_g (\chi_g) W_L(\chi_L, \chi_g) \\ \nonumber
\end{eqnarray}
with $P(k)$ being the mass power spectrum, and $J_2$ is the second
order Bessel function:
\begin{eqnarray}
J_n (y) = {1\over 2\pi} \int_{-\pi}^\pi d\zeta {\,\rm cos\,}
[y {\,\rm sin\,} \zeta - n \zeta]
\end{eqnarray}
Note that $W_g(\chi_g)$ is the selection function defined in eq. 
(\ref{epsilonproj}) and (\ref{deltagproj2}), and $W_L(\chi_L)$ is
the lensing efficiency function as given in eq. (\ref{kappa}). 

We have separated the density-ellipticity correlation into 
two terms: Term A $\langle \delta_g \epsilon_t^{\rm ex.} \rangle_A$
arises from magnification bias; 
term B $\langle \delta_g \epsilon_t^{\rm ex.} \rangle_B$
arises even if magnification bias is absent. The latter is
very much analogous to the density-ellipticity correlation
commonly measured in galaxy-galaxy lensing, except that here
we do not measure the projected density from one population
of galaxies and the ellipticity from another -- instead, we only have a 
single galaxy distribution $W_g (\chi_g)$, from which both density 
and ellipticity is inferred. Note that term B vanishes
if $W_g (\chi_g)$ has zero width -- it is a non-zero width that
allows galaxies closer to us to lens galaxies further away from us, both
sets being drawn from the same $W_g (\chi_g)$. 

\begin{figure}[htb]
\centerline{\epsfxsize=9cm\epsffile{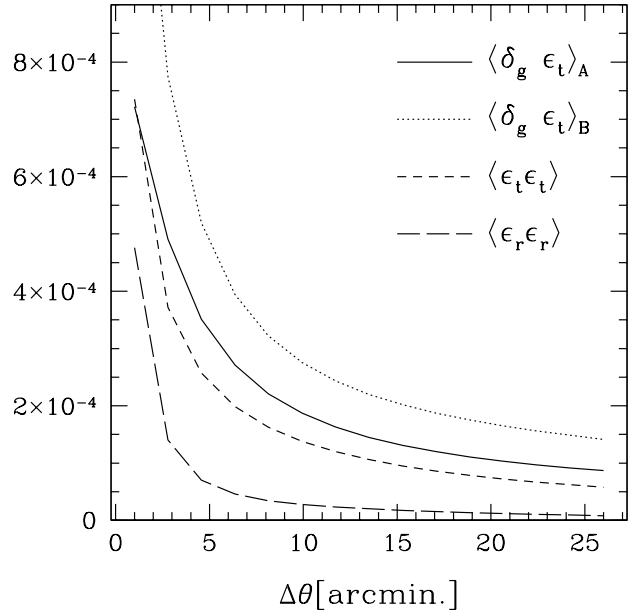}}
\caption{The lensing density-ellipticity correlation ($A$ and $B$, the
magnification bias term and the 'galaxy-galaxy lensing' term respectively,
see eq. [\ref{exdeltagepsilon}]), and the lensing ellipticity-ellipticity
correlations (eq. [\ref{epep}]), for a high redshift sample of source
galaxies (see eq. [\ref{Wg1}] for the selection function).}
\label{DE.paper}
\end{figure}

As an illustration, Fig. \ref{DE.paper} shows the two contributions to
$\langle \delta_g \epsilon_t \rangle$ as a function of angular
separation $\Delta \theta$ (solid and dotted lines) for 
a $\Lambda$CDM cosmological model ($\Omega_m = 0.3$, $\Omega_\Lambda = 0.7$, 
$\Gamma = 0.21$, $\sigma_8 = 0.9$), with the following selection function:
\begin{eqnarray}
\label{Wg1}
W_g (\chi_g) = N z^2 {\,\rm exp\,}[-(z/0.8)^{1.5}] |dz/d\chi_g | 
\end{eqnarray}
where $N$ is a normalizing factor chosen such that 
$\int d\chi_g W_g$ $(\chi_g) = 1$. Also shown in the figure are
the two components of the ellipticity-ellipticity lensing correlation
(short and long-dashed lines):
\begin{eqnarray}
\label{epep}
&& \langle \epsilon_t^{\rm ex.} (\bi{\theta}) \epsilon_t^{\rm ex.} (\bi{\theta'}) \rangle
= 4 \int {d\chi_L \over a^2} {\tilde W_L (\chi_L)^2 \over r(\chi_L)^2}
\\ \nonumber 
&& \quad \int_0^\infty {\ell d\ell \over 2 \pi} P(k = \ell/r(\chi_L))
{1\over 2} [J_0(\ell \Delta\theta) + J_4 (\ell \Delta\theta)] \\ \nonumber
&& \langle \epsilon_r^{\rm ex.} (\bi{\theta}) \epsilon_r^{\rm ex.} (\bi{\theta'}) \rangle
= 4 \int {d\chi_L \over a^2} {\tilde W_L (\chi_L)^2 \over r(\chi_L)^2}
\\ \nonumber 
&& \quad \int_0^\infty {\ell d\ell \over 2 \pi} P(k = \ell/r(\chi_L))
{1\over 2} [J_0(\ell \Delta\theta) - J_4 (\ell \Delta\theta)] 
\end{eqnarray}
(Kaiser 1992; Jain \& Seljak 1997). 
The power spectrum used is the nonlinear power spectrum
appropriate for the $\Lambda$CDM model 
(Hamilton et al. 1991, Jain, Mo \& White 1995, Peacock \& Dodds 1996). 
The model and selection function chosen here is
in rough accord with existing ellipticity-ellipticity measurements
from deep lensing surveys (e.g. van Waerbeke et al. 2000). 
The bias parameter and the slope of the luminosity function 
are chosen to be $b=1$ and $s=1$, for simplicity. 
If other values were chosen, it is simple to rescale our results:
$\langle \delta_g \epsilon_t \rangle_A \rightarrow
\langle \delta_g \epsilon_t \rangle_A \times (s - 0.4)/ 0.6$ and 
$\langle \delta_g \epsilon_t \rangle_B \rightarrow
\langle \delta_g \epsilon_t \rangle_B \times b$
(more generally, if $b$ and $s$ were redshift dependent,
that dependence has to be explicitly integrated; see
eq. [\ref{WLs}]). Note that if $s < 0.4$, the sign of the the $A$ term
is flipped. 
We should also note that on sufficiently small scales e.g. $\Delta\theta$
less than a few arcminutes, the assumption of a linear bias likely breaks down;
what is shown below nonetheless offers a rough estimate of the size of the 
signal on such scales.

\begin{figure}[htb]
\centerline{\epsfxsize=9cm\epsffile{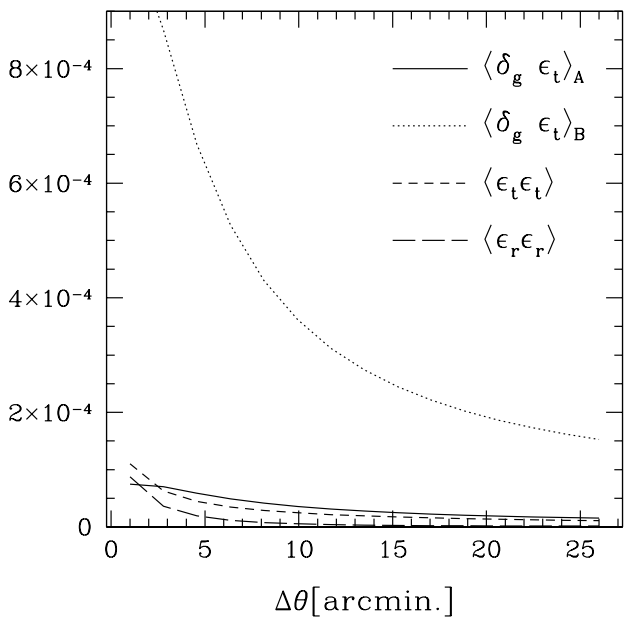}}
\caption{Same as Fig. \ref{DE.paper} except that the selection function is that
of a shallower survey (eq. [\ref{Wg2}]).}
\label{DE.paper.lowz}
\end{figure}

From Fig. \ref{DE.paper}, one can see that the $A$ and $B$ contributions to 
the lensing density-ellipticity correlation are comparable, which are
also similar in magnitude to the ellipticity-ellipticity correlations. 
Such a conclusion, however, is sensitive to the selection function 
of one's survey. 
Fig. \ref{DE.paper.lowz} shows the same set of quantities for
a source sample with lower redshifts:
\begin{eqnarray}
\label{Wg2}
W_g \propto z^2 {\,\rm exp \,} [-(z/0.35)^{1.7}] |dz/d\chi_g |
\end{eqnarray}
Lower source redshifts decrease the lensing efficiency $W_L$ (or $\tilde W_L$; eq. [\ref{WLs}]),
which lead to a drop in all the lensing correlations, except the $B$ term for
the density-ellipticity correlation. This is the term that arises from galaxy-galaxy
lensing within the sample. Two opposing effects roughly cancel out each other in this
case: on the one hand, lower source redshifts lead to less efficient lensing;
on the other hand, a shallower survey helps prevent projection from washing out
the galaxy density fluctuation $\delta_g$. 

\begin{figure}[htb]
\centerline{\epsfxsize=9cm\epsffile{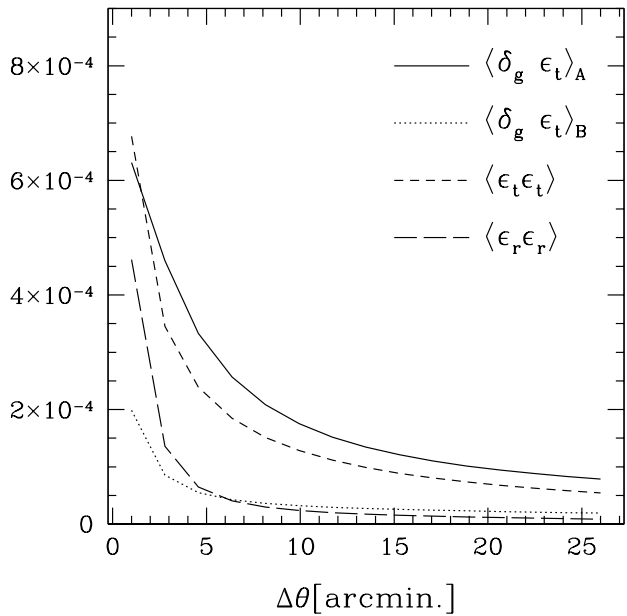}}
\caption{Same as Fig. \ref{DE.paper} except that the selection function is
chosen to be much more narrow (eq. [\ref{Wg3}]).}
\label{DE.paper.narrow}
\end{figure}

The 'galaxy-galaxy lensing' term ($B$ term) can be made much smaller if we
choose a selection function that is sufficiently narrow.
This is illustrated in Fig. \ref{DE.paper.narrow} which employs the following
selection function
\begin{eqnarray}
\label{Wg3}
W_g \propto {\,\rm exp\,}[- (z - 1)^2 / 2 / 0.05^2 ] |dz/d\chi_g |
\end{eqnarray}
Here, the $B$ term is about a factor of 5 smaller than the $A$ term, in
contrast with the previous two examples. The important point to keep in
mind is that the magnification bias term ($A$ term) and
the ellipticity correlation functions are not sensitive to the 
width of the selection function, because for these quantities, $W_g$ enters only through
convolution with a rather broad lensing efficiency function (eq. [\ref{WLs}]). 
The $B$ term, on the other hand, can be made as small as one wishes if one has
the ability to measure redshifts accurately.

\section{What about non-Gaussianity?}
\label{nonGauss}

Are we to conclude from \S \ref{intrinsic} that the intrinsic density-ellipticity
correlation should strictly vanish? 
A crucial assumption made in the calculation of the intrinsic density-ellipticity
correlation in \S \ref{intrinsic} is that the fluctuation in gravitational
potential $\phi$ is Gaussian random.
While this is approximately true on large scales, it cannot be exact -- gravitational
instability induces non-Gaussianity from initially Gaussian conditions, even on large scales. 
It is important that we estimate the strength of the non-Gaussian corrections.
To do so properly, we have to make a digression and discuss ellipticity-ellipticity
correlation. Since non-Gaussian fluctuations are easier to include in the
framework of MWK, that is what we are going to adopt. 

In \S \ref{b2bias}, we will also briefly examine the consequence of allowing
for a nonlinear galaxy-bias. It turns out doing so brings in terms quite similar to
those introduced by non-Gaussianity, as far as the density-ellipticity correlation
is concerned. Because of the wider relevance of non-Gaussian terms (for
ellipticity-ellipticity as well as density-ellipticity correlations), they will take
up most of this section.

\subsection{A Digression on Intrinsic Ellipticity-Ellipticity Correlation}
\label{digression}

We take as starting point the expression for $\epsilon_i^{\rm in.}$ given in
eq. (\ref{mackey}), except that we now allow $\phi$ to be non-Gaussian in such
a way consistent with gravitational instability. 
The usual Gaussian assumption follows from working out the angular momentum
of galaxies due to tidal torque-up according to first order Lagrangian perturbation theory 
(Doroshkevich 1970). A higher order calculation\footnote{See Peebles (1969) and White (1984) for
second order calculations that assume a spherical Eulerian/Lagrangian volume.} 
would obviously
introduce non-Gaussian terms, but it should be emphasized that {\it a consistent expansion
does not necessarily yield an expression for $\epsilon_i^{\rm in.}$ that is
given exactly by eq. (\ref{mackey}) where $\phi$ is simply replaced by its non-linear
counterpart.}
Nonetheless, this should provide a useful order of magnitude estimate.
It can be shown that a second order expansion does give rise to terms similar
to the ones we consider here, albeit with different coefficients of the same order.
Further details will be given in a separate paper (Zhang \& Hui 2002). 

Because we are interested in the relative significance of non-Gaussian versus Gaussian terms,
some care must be taken in choosing the time at which the spins of galaxies 
(and therefore the tidal fields) are to be evaluated. 
According to the usual tidal torque-up theories, galaxies or halos gain most of their
angular momentum before turn-around, after which the intrinsic correlations
are presumably locked in place. However, as Porciani et al. (2001a,b) pointed out
(see also Sugerman et al. 2000),
random kicks by accreting satellites cause a misalignment of the eventual angular momentum
from the one predicted by tidal torque-up theories (van den Bosch et al. 2002 also showed
that hydrodynamic effects cause a misalignment between the angular momentum of the gas
with respect to that of the dark matter). The hope is that a reduction
of the intrinsic correlations due to such effects 
occurs in a scale-independent way, and can therefore
be taken care of by rescaling the overall amplitude of the intrinsic correlations
(as is done by MWK). Porciani et al. (2001a) showed that angular momentum growth slows
down after $z \sim 3$, while misalignment starts to grow quite early on, from $z \sim 50$
onward. In what follows, to illustrate the range of possibilities, 
we will therefore consider two cases: the tidal field $\phi$ in 
the expressions for the intrinsic ellipticities 
(eq. [\ref{mackey}]) are to be evaluated at
$z_{\rm T} = 3$ or $z_{\rm T} = 50$. 
\footnote{The mapping from Lagrangian to Eulerian space which moves the galaxies around
also causes an evolution of the intrinsic correlations. A glance at Fig. 11 of
Porciani et al. (2001a) suggests this is a small effect.}

Let us focus on $\langle \epsilon_2^{\rm in.} (\bi{\theta}) \epsilon_2^{\rm in.} 
(\bi{\theta'}) \rangle$ for an order of magnitude estimate:
\begin{eqnarray}
\label{nonGstart}
&& \langle \epsilon_2^{\rm in.} (\bi{\theta}) \epsilon_2^{\rm in.} 
(\bi{\theta'}) \rangle = \int d\chi_g W(\chi_g)^2 \int dp_3 
{\beta^2 \over 15^2} \\ \nonumber &&
\langle 
[4 \phi_{33} \phi_{12} - 2 \phi_{11} \phi_{12} - 2 \phi_{22} \phi_{12}
- 6 \phi_{13} \phi_{23}]_{\bf p} \\ \nonumber &&
[4 \phi_{33} \phi_{12} - 2 \phi_{11} \phi_{12} - 2 \phi_{22} \phi_{12}
- 6 \phi_{13} \phi_{23}]_{\bf q} \rangle
\end{eqnarray}
where we have employed Limber's approximation and used the same notation
as in eq. (\ref{Limber}): ${\bf p_\perp} - {\bf q_\perp} 
= r(\chi_g) (\bi{\theta}-\bi{\theta'})$; $\beta$ is as given in eq. (\ref{mackey}). 

Let us consider for example the term:
\begin{eqnarray}
\label{oneterm}
&& \langle \phi_{33} ({\bf p}) \phi_{12} ({\bf p}) \phi_{33} ({\bf q}) 
\phi_{12} ({\bf q}) \rangle 
= \\ \nonumber && 
\langle \phi_{33} ({\bf p}) \phi_{12} ({\bf p}) \rangle
\langle \phi_{33} ({\bf q}) \phi_{12} ({\bf q}) \rangle
\\ \nonumber &&
+ \langle \phi_{33} ({\bf p}) \phi_{33} ({\bf q}) \rangle
\langle \phi_{12} ({\bf p}) \phi_{12} ({\bf q}) \rangle
\\ \nonumber &&
+ \langle \phi_{33} ({\bf p}) \phi_{12} ({\bf q}) \rangle
\langle \phi_{33} ({\bf p}) \phi_{12} ({\bf q}) \rangle
\\ \nonumber &&
+ \langle \phi_{33} ({\bf p}) \phi_{12} ({\bf p}) \phi_{33} ({\bf q}) 
\phi_{12} ({\bf q}) \rangle_c
\end{eqnarray}
The first term on the right can be ignored due to isotropy.
The next two terms are Gaussian terms which depend quadratically on
the power spectrum, and are included in the analysis of MWK. 
The last term on the right is the non-Gaussian term.
One can write the above as:
\begin{eqnarray}
&& 
(4\pi G \bar\rho a^2)^{-4}
\langle \phi_{33} ({\bf p}) \phi_{12} ({\bf p}) \phi_{33} ({\bf q}) 
\phi_{12} ({\bf q}) \rangle = \\ \nonumber 
&& \int {d^3 k^A \over (2\pi)^3} P(k^A) {(k^A_3)^4 \over (k^A)^4}
e^{-i {\bf k^A} \cdot ({\bf p} - {\bf q})} \\ \nonumber 
&& \int {d^3 k^B \over (2\pi)^3} P(k^B) {(k^B_1 k^B_2)^2 \over (k^B)^4}
e^{-i {\bf k^B} \cdot ({\bf p} - {\bf q})} \\ \nonumber
&& + \left[\int {d^3 k^A \over (2\pi)^3} P(k^A) {(k^A_3)^2 k^A_1 k^A_2 \over (k^A)^4}
e^{-i {\bf k^A} \cdot ({\bf p} - {\bf q})}  \right]^2 \\ \nonumber 
&& + \int {d^3 k^A \over (2\pi)^3} {d^3 k^B \over (2\pi)^3} 
{d^3 k^C \over (2\pi)^3} T({\bf k^A, k^B, k^C, k^D}) \\ \nonumber &&
{(k^A_3)^2 \over (k^A)^2} {k^B_1 k^B_2 \over (k^B)^2} 
{(k^C_3)^2 \over (k^C)^2} {k^D_1 k^D_2 \over (k^D)^2} 
e^{-i ({\bf k^A} \cdot {\bf p} + {\bf k^B} \cdot {\bf p}
+ {\bf k^C} \cdot {\bf q} + {\bf k^D} \cdot {\bf q})}
\end{eqnarray}
where ${\bf k^A} + {\bf k^B} + {\bf k^C} + {\bf k^D} = 0$, and
$T$ is the trispectrum. Here, the power spectrum $P$ and the trispectrum
refers to those of the mass $\delta$, and the factor of 
$(4\pi G \bar\rho a^2)^{-4}$ (together with factors of $k^2$) takes care of the scaling between
$\phi$ and $\delta$. 

The trispectrum $T$ scales like the power spectrum cubed, a scaling
motivated by perturbation theory (under Gaussian initial conditions) 
and confirmed by numerical N-body
simulations even in the nonlinear regime (see Fry 1984, Scoccimarro et al. 1998, 
Baugh et al. 1995, but see also Suto \& Matsubara 1994, Jing \& Boerner 1998):
\begin{eqnarray}
&& T({\bf k^A, k^B, k^C, k^D}) = \\ \nonumber 
&& R_a [ P(k^A) P({\bf k^A + k^B})
P(k^C) + 
{\,\rm 11 \, other \, cyclic \, perm.} ] \\ \nonumber 
&& + R_b [P(k^A) P(k^B) P(k^C) + 
{\,\rm 3 \, other \, cyclic \, perm.}]
\end{eqnarray}
The coefficients $R_a$ and $R_b$ are configuration dependent on large
scales, but asymptotes to constants on small scales. Both are
expected to vary somewhat slowly with scale when averaged over
configurations (e.g. see variation of
the kurtosis, which is related to $R_a$ and $R_b$ in an averaged sense,
as a function of scale in Baugh et al. 1995). As a rough order of magnitude 
estimate,
we will follow Scoccimarro et al. (1999) to treat $R_a \sim R_b \sim Q_4$,
where $Q_4$ is given by Hyper-Extended Perturbation Theory (HEPT; Scoccimarro \& Frieman 1999).
We will use $Q_4 \sim 3$. One should keep in mind that HEPT
strictly applies only at highly nonlinear scales, and likely leads to some overestimate
of the trispectrum term, but is probably acceptable in the spirit of an order
of magnitude estimate.
\footnote{Ours is a particularly demanding application of HEPT, because 
we are evaluating the four-point function at two pairs of points where
each pair consists of closely separated points, while the two pairs are
widely separated from each other.}

Putting everything together, one obtains the following
approximate form for eq. (\ref{nonGstart}):
\begin{eqnarray}
\label{nonGstart2}
&& \langle \epsilon_2^{\rm in.} (\bi{\theta}) \epsilon_2^{\rm in.} 
(\bi{\theta'}) \rangle \sim [\chi_* \int d\chi_g W(\chi_g)^2] \\ \nonumber &&
A^2 {\beta^2 \over 15^2} {r(\chi_*) \Delta\theta \over \chi_*} (3\Omega_m H_0^2/2)^4
\\ \nonumber && (1+z_{\rm T})^4 [ \pi  \Delta^4(k_*) + 
{\pi} Q_4 
\Delta^2 (k_*) \Delta^4 (k_R) 
\\ \nonumber &&
+ 2 {\pi} Q_4 \Delta^4 (k_*) \Delta^2 (k_R) + \pi Q_4 \Delta^6 (k_*)]_{z_{\rm T}}
\end{eqnarray}
where $\chi_*$ is the typical (comoving) radial distance of the galaxies,
$k_* \equiv 1/r(\chi_*)/\Delta\theta$, $k_R \equiv 1/R$ where $R$
is the appropriate smoothing scale for the galaxies (taken to be $1$ Mpc/h by MWK), and
$\Delta^2 (k) \equiv 4\pi k^3 P(k)/(2\pi)^3$, with $P(k)$
evaluated at the redshift $z_{\rm T}$ 
(which will be taken to be $3$ or $50$, higher than
the typical redshift of the galaxies which we will refer to as $z_*$ 
(the latter corresponds to $\chi_*$)). 
\footnote{The expression in eq. (\ref{nonGstart2}), 
even if one ignores the non-Gaussian terms, 
differs slightly from the corresponding one given by MWK in its redshift ($z_*$) 
dependence. Essentially, MWK sets $z_{\rm T} = z_*$. To the extent that
$(1+z)^4 \Delta^4 (k, z)$ is only weakly redshift dependent, this difference
is small.}
We have used the Poisson equation here to relate mass fluctuation $\delta$ and $\phi$ 
(eq. [\ref{linearbias}]), hence the factors of $3H_0^2 \Omega_m / 2$.
It should be noted that the integrals over multiple power spectra leading to terms like 
$\Delta^4 (k_*)$ or $\Delta^2 (k_*) \Delta^4 (k_R)$ do not strictly factorize, and
eq. (\ref{nonGstart2}) should be viewed as a rough approximation. 
To be concrete: for instance, one of the Gaussian terms goes like 
\begin{eqnarray}
\nonumber
\int {d^3 k^A \over (2\pi)^3} {d^3 k^B \over (2\pi)^3}
 P(k^A) P(k^B) 2\pi \delta (k^A_3 + k^B_3) \\ \nonumber G(\hat {\bf k^A}, \hat {\bf k^B}) 
e^{-i ({\bf k^A + k^B}) \cdot ({\bf p - q})}
\end{eqnarray}
where the delta function comes from the Limber's approximation integral over
$dp_3$ (eq. [\ref{nonGstart}]), and $G$ represents some function of the unit
vectors $\hat {\bf k^A} \equiv {\bf k^A}/k^A$ and $\hat {\bf k^B} \equiv {\bf k^B}/k^B$.
We approximate the above as $\propto \Delta^4 (k_*) / k_*$, where
$k_* = 1/({\bf p-q})_\perp = 1/\Delta\theta/r(\chi_*)$. 

For the normalization factor $\beta$, MWK recommended using 
\begin{eqnarray}
\label{beta}
\beta = \sqrt{1125 \over 32} {0.4 \over 2.8^2} {1 \over (3\Omega_m H_0^2/2)^2}
\end{eqnarray}
We have introduced an additional overall factor of $A^2$ in eq. (\ref{nonGstart2}), because 
we need to rescale our result so that the correlation still has the desired
amplitude after the non-Gaussian terms are introduced (also because we have chosen to
evaluate the tidal field correlations at ${z_T}$). 
We adopt the choice that $\langle \epsilon_2^{\rm in.} (\bi{\theta}) \epsilon_2^{\rm in.} 
(\bi{\theta'}) \rangle$ agrees roughly with MWK today at $k_* = k_R$ i.e. we choose
\begin{eqnarray}
\label{Afact}
A^2 = {\pi \Delta^4 (k_R, z=0) \over 
(1+z_{\rm T})^4 [\pi \Delta^4 (k_R) + 4 \pi Q_4 \Delta^6 (k_R)]_{z = z_{\rm T}}}
\end{eqnarray}
\footnote{It should be noted, as MWK pointed out, the formulation laid down in
eq. (\ref{mackey}) for the relation between the intrinsic ellipticity and the tidal field
is, strictly speaking, a little problematic, since it allows the possibility of 
$\epsilon_i^{\rm in.} > 1$.
They found that in practice the probability of that happening remains small as long as
they choose $\beta$ properly. The tuning of $A$ does the same thing for us here.
But one should keep in mind a non-Gaussian tail probably makes the problem a little worse.}

A crude way to understand our result in eq. (\ref{nonGstart2}) is this.
Think of $\epsilon_i^{\rm in.} \propto \phi^2$ (ignoring derivatives). 
Therefore we have 
$\langle \epsilon_i^{\rm in.} ({\bf p}) \epsilon_i^{\rm in.} ({\bf q}) \rangle$
very roughly given by
\begin{eqnarray}
\label{crude}
& \langle \phi^2 ({\bf p}) \phi^2 ({\bf q}) \rangle 
\sim \langle \phi ({\bf p}) \phi ({\bf q}) \rangle^2 + \\ \nonumber &
\langle \phi ({\bf p}) \phi ({\bf q}) \rangle \langle \phi^2 \rangle
\langle \phi^2 \rangle + \\ \nonumber &
\langle \phi ({\bf p}) \phi ({\bf q}) \rangle^2 \langle \phi^2 \rangle
+ \langle \phi ({\bf p}) \phi ({\bf q}) \rangle^3
\end{eqnarray}
where we have not been careful about the coefficients of each term.
We have also ignored a term that goes like $\langle \phi^2 \rangle^2$, which
if worked out properly with the right combination of derivatives, gives
$\langle \epsilon_i^{\rm in.} \rangle^2 = 0$. 
The first term on the right is the usual Gaussian result, giving a quadratic scaling with 
the two-point function. The second, third and fourth terms arise from non-Gaussianity (using
the hierarchical scaling of the trispectrum), and the second term in particular 
gives us a linear scaling with the two-point
function. It should be kept in mind that the above factorization is actually
not exact once derivatives are taken into account. One might also worry that
the factors of $\langle \phi^2 \rangle$ might yield zero once the various
derivatives are properly combined. It can be shown that this does not happen.

\begin{figure}[htb]
\centerline{\epsfxsize=9cm\epsffile{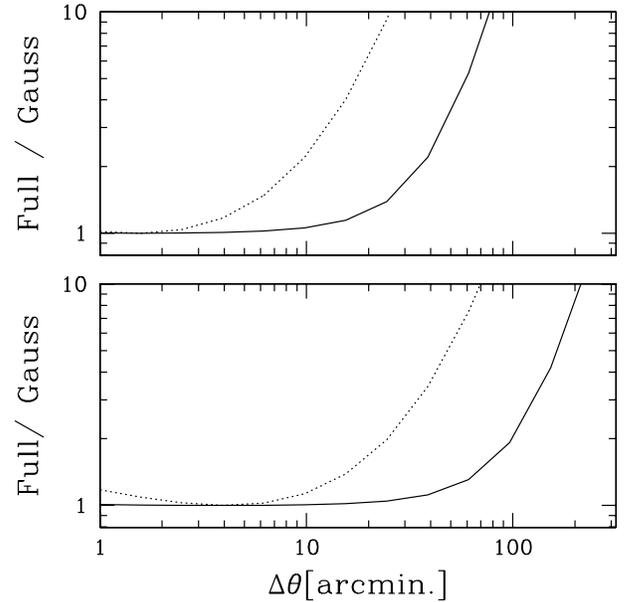}}
\caption{Ratio of the full expression (including non-Gaussian terms; eq. [\ref{nonGstart2}]) for
the intrinsic ellipticity-ellipticity correlation to its Gaussian version (i.e.
setting $Q_4 = 0$ in both eq. [\ref{nonGstart2}] and [\ref{Afact}]). 
The upper panel assumes a selection function appropriate for a deep survey 
(eq. [\ref{Wg1}]), while the lower panel uses that for a shallow survey (eq. [\ref{Wg2}]). 
Dotted lines denote the ratio when using $z_{\rm T} = 3$, while solid lines
use $z_{\rm T} = 50$.}
\label{epepratio}
\end{figure}

The key implication of eq. (\ref{crude}) (and eq. [\ref{nonGstart2}]) is that 
at least one of the non-Gaussian terms is non-negligible on sufficiently large
scales. {\it At a sufficiently large angular separation, the non-Gaussian term that scales
linearly with the two-point function or power spectrum {\rm ($\propto \Delta^2 (k_*)$)}
will dominate over the usual Gaussian term {\rm ($\propto \Delta^4 (k_*)$)}.}
More precisely, on sufficiently large scales, the dominant term scales
with the angular separation as $\Delta\theta \Delta^2 (k_* \propto 1/\Delta\theta)$. 
At what scale this happens depends on the depth of the survey as well as
on $z_{\rm T}$, the redshift at which the tidal correlations are to be evaluated. 
Examples are shown in Fig. \ref{epepratio}. 
We use the linear power spectrum corresponding to 
the $\Lambda$CDM model discussed in \S \ref{extrinsic} when evaluating eq. (\ref{nonGstart2})
(and similarly for the rest of this section). 

The results in Fig. \ref{epepratio} might seem a little surprising, especially
for the case where $z_{\rm T} = 50$, so that $\Delta (k_R) = 0.056$
(recall that $\Delta (k)$ is evaluated at $z_{\rm T}$), which may
lead one to think that non-Gaussianity can be completely ignored. 
This is true of the normalizing factor (eq. [\ref{Afact}]), where one
can simply ignore the term proportional to $Q_4$. However, the same cannot be
done for eq. (\ref{nonGstart2}). At sufficiently large scales (small $k_*$), 
$\Delta (k*)$ is small enough such that $\Delta^4 (k_*) < Q_4 \Delta^2 (k*) \Delta^4 (k_R)$.
\footnote{One might wonder if the significance of the non-Gaussian terms can be
reduced if the ellipticity is smoothed over some large scale by hand. This corresponds
to smoothing $\epsilon_i^{\rm in.} \sim \phi^2$. Because such an 'after-the-fact' smoothing is not
applied to $\phi$ but to $\phi^2$, it can be shown that the significance of
the non-Gaussian terms compared to the Gaussian one cannot be reduced in this way.
Therefore, as long as $\phi$ can be viewed as being smoothed on a small (galactic) scale
$k_R \sim 1$ h/Mpc, our conclusions remain valid.
}
\footnote{We note that in the case of $z_{\rm T} = 3$, perturbation theory
is close to breaking down, because $\Delta (k_R)$ approaches unity.}

Our argument here is actually quite similar to the well-known derivation of linear biasing in
the case of the galaxy correlation. Suppose 
$\delta_g$ is some local nonlinear function of $\delta$, one can argue on quite
general grounds that $\langle \delta_g ({\bf p}) \delta_g ({\bf q}) \rangle$
scales linearly with $\langle \delta ({\bf p}) \delta ({\bf q}) \rangle$ for
sufficiently large $|{\bf p} - {\bf q}|$
(Scherrer \& Weinberg 1998).

Since the extrinsic correlation is also expected to 
scale linearly with the power spectrum, our calculation implies that the ratio of the
intrinsic signal to extrinsic signal does not actually drop with scale (on large scales), unlike
the case of MWK or CNPT who considered the Gaussian term only. 
This is reminiscent of the conclusion drawn by
Catelan et al. (2001), although the reason is quite different -- we assume
here that the galaxy ellipticity depends quadratically on the gravitational potential 
$\phi$ (following MWK and CNPT), while Catelan et al. employed a linear relation. 

\subsection{Non-Gaussian Intrinsic Density-Ellipticity Correlation}
\label{nongaussDE}

Using very similar arguments to the above, we can write down the following
approximate expression for the intrinsic density-ellipticity correlation:
\begin{eqnarray}
\label{nongaussDEin}
&& \langle \delta_g (\bi{\theta}) \epsilon_t^{\rm in.} (\bi{\theta'}) \rangle
\sim [\chi_* \int d\chi_g W_g (\chi_g)^2 ] {\beta A \over 15} 
\\ \nonumber && (3\Omega_m H_0^2/2)^2 {D(z_*) \over D(z_{\rm T})} {\pi \over 2} Q_3 {r(\chi_*) 
\Delta\theta \over \chi_*} b
\\ \nonumber && (1+z_{\rm T})^2 [\Delta^4 (k_*) + 2 \Delta^2 (k_*) \Delta^2 (k_R)]_{z_{\rm T}}
\end{eqnarray}
where we have used the hierarchical scaling: the mass bispectrum $B({\bf k^A}, {\bf k^B}, {\bf k^C})
= Q_3 [ P(k^A) P(k^B) + P(k^B) P(k^C) + P(k^C) P(k^A) ]$, with $Q_3 \sim 2$ (
using HEPT as before; Scoccimarro \& Frieman 1999). We have ignored the source-clustering
correction here.\footnote{Source-clustering corrections in principle can introduce terms
that are important as well, but they will not alter greatly our order of magnitude estimates,
and will not change our conclusion regarding the scaling of the correlations. We therefore
ignore them for simplicity.}
The $b$ here is the linear bias factor (eq. [\ref{linearbias}]).
The factor $D(z_*)/D(z_{\rm T})$ stands for the ratio of the linear
growth between $z_*$ and $z_{\rm T}$. This accounts for the linear growth
of $\delta_g$ (or $\delta$) between the two different redshifts. 

\begin{figure}[htb]
\centerline{\epsfxsize=9cm\epsffile{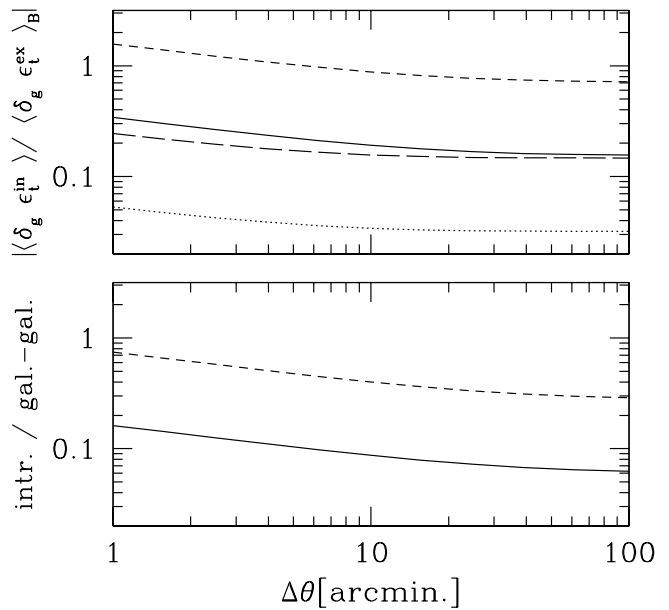}}
\caption{Ratio of intrinsic signal to extrinsic signal. 
{\it Upper panel:} the ratio of eq. (\ref{nongaussDEin}) to eq. (\ref{DEex}) for
the density-ellipticity correlation, for two different selection functions.
The short-dashed and solid lines correspond to a shallow survey 
($W_g$ given by eq. [\ref{Wg2}]), with the short-dashed line using $z_{\rm T} = 3$,
and the solid line using $z_{\rm T} = 50$. 
The long-dashed and dotted lines correspond to a deep survey
($W_g$ given by eq. [\ref{Wg1}]), with the long-dashed line using $z_{\rm T} = 3$,
and the dotted line using $z_{\rm T} = 50$. 
{\it Lower panel:} the level of contamination
from intrinsic alignment to a galaxy-galaxy lensing measurement of 
density-ellipticity correlation (ratio of eq. [\ref{nongaussDEin2}] to 
[\ref{DEex2}]), using two different prescriptions for $z_{\rm T}$. 
The dashed line uses $z_{\rm T} = 3$ and the solid line uses $z_{\rm T} = 50$. 
The foreground and background selection functions here are
chosen to mimic the SDSS galaxy-galaxy lensing survey of Fischer et al. (2000).
}
\label{ratio}
\end{figure}

It is interesting to compare this against the extrinsic density-ellipticity correlation,
e.g. the B-term (eq. [\ref{exdeltagepsilon}]):
\begin{eqnarray}
\label{DEex}
\langle \delta_g (\bi{\theta}) \epsilon_t^{\rm ex.} \rangle_B \sim
3 b \Omega_m (1+z_*) 
\pi r(\chi_*)^2 H_0^2 \Delta\theta \Delta^2(k_*) \\ \nonumber 
[\int d\chi_L W_g (\chi_L) \int_{\chi_L}^\infty W_g (\chi_g) r(\chi_g - \chi_L)/r(\chi_g)]
\end{eqnarray}
where $\Delta^2 (k_*)$ is evaluated at $z_*$, the typical redshift of the galaxies. 
The ratio between the intrinsic and extrinsic density-ellipticity correlations
is shown in the upper panel of Fig. \ref{ratio}, for
a deep survey ($z_* \sim 1$, $W_g$ as given by eq. [\ref{Wg1}]; 
long-dashed line for $z_{\rm T} = 3$, and dotted  line for $z_{\rm T} = 50$, where
$z_{\rm T}$ is the redshift at which tidal correlations are evaluated), and a shallow survey 
($z_* \sim 0.3$, $W_g$ given by eq. [\ref{Wg2}]; short-dashed line for $z_{\rm T} = 3$, 
and solid line for $z_{\rm T} = 50$). 
Intrinsic alignment induces a non-negligible contribution to the observed
density-ellipticity correlation, especially for a shallow survey.
For a sufficiently deep survey, however, 
intrinsic alignment contributes only at the level of a few percent, if the 
redshift at which tidal correlations are evaluated, $z_{\rm T}$, is large. 
The latter suggests that a measurement of both
ellipticity-ellipticity and density-ellipticity correlations from a deep survey
might offer a nice consistency check of the lensing hypothesis, provided the
galaxy bias can be independently constrained. 
We will come back to this point in
\S \ref{discuss}. 

We caution that our estimates in Fig. \ref{ratio} on small angular scales
(less than several arcminutes) should be viewed with some skepticism, 
since $\delta$ (or $\delta_g$) has likely gone nonlinear in this regime.

\subsection{Application to Galaxy-Galaxy Lensing}
\label{galgal}

The finding that the intrinsic density-ellipticity correlation is generally non-zero
naturally raises the question of how this might impact measurements of galaxy-galaxy lensing.
Typically foreground and background populations of galaxies are defined using
a simple magnitude cut or photometric redshifts (or in the case of McKay et al. 2001, using
spectroscopic redshifts to define foreground, but not the background). Generally, the
two populations will have a non-zero overlap, and the question is whether 
intrinsic alignment constitutes a significant contaminant through this overlap. 

The intrinsic density-ellipticity correlation for such a set-up is
very similar to that given in eq. (\ref{nongaussDEin}), except that
one of the $W_g$ is replaced by $W_g^f$, the foreground selection function
(the other $W_g$ is used to refer to the background):
\begin{eqnarray}
\label{nongaussDEin2}
&& \langle \delta_g (\bi{\theta}) \epsilon_t^{\rm in.} (\bi{\theta'}) \rangle
\sim [\chi_* \int d\chi_g W_g (\chi_g) W_g^f (\chi_g) ] \\ \nonumber &&
{\beta A \over 15} (3\Omega_m H_0^2/2)^2 {\pi \over 2} Q_3 {r(\chi_*) 
\Delta\theta \over \chi_*} b {D(z_*) \over D(z_{\rm T})}
\\ \nonumber && (1+z_{\rm T})^2 [\Delta^4 (k_*) + 2 \Delta^2 (k_*) \Delta^2 (k_R)]_{z_{\rm T}}
\end{eqnarray}

Similarly, the extrinsic density-ellipticity correlation (galaxy-galaxy lensing)
is given by the generalization of eq. (\ref{DEex}):
\begin{eqnarray}
\label{DEex2}
\langle \delta_g (\bi{\theta}) \epsilon_t^{\rm ex.} \rangle \sim
3 b \Omega_m (1+z_*) 
\pi r(\chi_*)^2 H_0^{-2} \Delta\theta \Delta^2(k_*) \\ \nonumber 
[\int d\chi_L W_g^f (\chi_L) \int_{\chi_L}^\infty W_g (\chi_g) r(\chi_g - \chi_L)/r(\chi_g)]
\end{eqnarray}
where $\Delta^2 (k_*)$ is evaluated at $z_*$. 

As an illustration, for the background selection function, we adopt $W_g$ given by eq. (\ref{Wg2});
for the foreground, we use
\begin{eqnarray}
W_g^f (\chi_g) = z^2 {\,\rm exp\,} [- (z/0.17)^{2.3} ] |dz/d\chi_g |
\end{eqnarray}
These parameters are supposed to describe the galaxy-galaxy lensing configuration
of Fischer et al. (2000) (see Guzik \& Seljak 2001). 
The ratio between eq. (\ref{nongaussDEin2}) and (\ref{DEex2}) is shown in the
lower panel of Fig. \ref{ratio}: solid line for $z_{\rm T} = 50$, and dashed line for $z_{\rm T} = 3$. 
Therefore, at a scale of 10 arcminutes, a conservative estimate of 
the intrinsic alignment contamination to galaxy-galaxy lensing is 
about $10 \%$. We will discuss uncertainties in this estimate in \S \ref{discuss}.

\subsection{Nonlinear Bias}
\label{b2bias}

We have assumed a linear galaxy-bias $\delta_g \propto \delta$ so far. 
On small scales, it is almost for certain that the relationship between $\delta_g$ and 
$\delta$ will be more complicated. Recent work using halo models provides perhaps
the most sophisticated description of the galaxy-mass relation (e.g. 
Peacock \& Smith 2000, Seljak 2000, Ma \& Fry 2000, Scoccimarro et al. 2001). 
It is in principle possible to use such models to predict the intrinsic density-ellipticity
correlation. This is more properly dealt with in a separate paper. Here we would like to simply
point out that a nonlinear bias brings in terms rather similar to
those we have considered. 

Suppose $\delta_g = b \delta + b_2 \delta^2$ (see Fry \& Gaztanaga 1993), where
$b_2$ is some constant. Then, $\langle \delta_g ({\bf p}) \epsilon^{\rm in.}_i ({\bf q}) \rangle$
contains a term that goes like $b_2 \langle \delta ({\bf p}) \delta ({\bf p}) \phi ({\bf q}) 
\phi ({\bf q}) \rangle$, where we have suppressed the derivatives on $\phi$. 
For a Gaussian random field ($\delta$ and $\phi$), this would produce terms
that scale quadratically with the two-point function. If $\delta$ and $\phi$ were non-Gaussian,
there can be terms that scale linearly with the two-point correlation as well.
It is also worth noting that on large scales, it is quite likely
$b_2$ is small (see e.g. Scoccimarro et al. 2001). 

\section{Discussion}
\label{discuss}

Our findings are summarized as follows:

{\bf 1.} According to two different tidal alignment theories
(Crittenden et al. 2001 [CNPT] \& Mackey et al. 2001 [MWK]), the 
intrinsic density-ellipticity correlation should vanish exactly. 
This includes both the lowest order term $\langle 
\delta_g (\bi{\theta}) \epsilon_i^{\rm in.} (\bi{\theta'}) \rangle$, 
as well as the source clustering correction 
$\langle \delta_g (\bi{\theta}) \epsilon_i^{\Delta\,\rm in.} (\bi{\theta'}) 
\rangle$, where $i = 1, 2$ or $i = t$
(that the $i=r$ component vanishes is guaranteed by parity invariance alone;
see eq. [\ref{epsilontr}]). 
This calculation makes two main assumptions: a Gaussian random gravitational
potential field (as is assumed by CNPT and MWK) and linear biasing
(eq. [\ref{linearbias}]). While linear biasing is a good approximation on
large scales, Gaussianity might not be a good one even on large scales (see below).

{\bf 2. } We have computed the extrinsic density-ellipticity correlation
due to weak gravitational lensing. Only the lowest order term 
$\langle \delta_g (\bi{\theta}) \epsilon_t^{\rm ex.} (\bi{\theta'}) \rangle$
is considered here (i.e. the source clustering correction is
ignored); it is non-zero in general. There are two 
contributions to it (eq. [\ref{exdeltagepsilon}): 
one arises from magnification bias (that 
gravitational lensing modulate the observed density of sources in a
flux limited survey; this we call the $A$-term), and the other
is analogous to what is commonly known as galaxy-galaxy lensing, except
that here we are interested in galaxies drawn from a single selection function
(foreground galaxies lensing background galaxies
from the same selection function of some finite width; this we call the
$B$-term). Illustrations of these two terms, compared against the
more familiar ellipticity-ellipticity lensing correlations, are
shown in Fig. \ref{DE.paper} to \ref{DE.paper.narrow}. Two main trends
are noteworthy. First, the $B$ term is systematically higher than
all other terms if the source galaxies are at sufficiently low redshifts
($z \lsim 0.3$); by $z \sim 1$, all terms are roughly comparable. 
Second, the $A$ term can be made dominant over the $B$ term if
the width of the selection function can be made sufficiently narrow. 

{\bf 3.} We have considered non-Gaussian contributions to both ellipticity-ellipticity
and density-ellipticity intrinsic correlations. They are non-negligible.

{\it a.} In the case of the intrinsic ellipticity-ellipticity correlation, the non-Gaussian
contributions predict a large scale linear scaling with the power spectrum
rather than a quadratic one as discussed by MWK and CNPT.
The result is somewhat surprising, in that higher order terms actually
dominate over lower order ones (see \S \ref{digression}). 
It originates from the fact that the lowest order Gaussian term scales with
the angular separation like $\Delta\theta \Delta^4 (k_*)$ while one of the higher order non-Gaussian
terms scales as $\Delta\theta \Delta^2 (k_*)$, where
$\Delta(k_*)$ is the fluctuation amplitude at wavenumber $k_* \propto 1/\Delta\theta$
(eq. [\ref{nonGstart2}]). Even though the higher order term is suppressed by a small
coefficient, it still dominates at sufficiently large angles $\Delta\theta$.
The amplitude of the resulting ellipticity-ellipticity correlation is somewhat uncertain -- 
we fix it by matching the results of MWK on small scales (so as to obtain the right
rms value for the observed ellipticities); but the large scale scaling we believe to be robust.

{\it b.} We compute the intrinsic density-ellipticity correlation, which 
receive contributions from non-Gaussian terms alone. A comparison of this with
the extrinsic density-ellipticity correlation is shown in Fig. \ref{ratio}. 
For a shallow survey (e.g. median redshift of $0.3$), the intrinsic signal is non-negligible.
However, for a sufficiently deep survey (i.e. median redshift of $1$ or above), 
the intrinsic signal is only a small fraction (a few percent) of the extrinsic signal
on large scales, if $z_{\rm T}$, the redshift at which tidal correlations are
evaluated, is large enough. 

{\it c.} We apply the calculation of density-ellipticity correlation to
the case of galaxy-galaxy lensing, and find that for a low redshift survey such
as SDSS (Fischer et al. 2000), intrinsic alignment constitutes a non-negligible 
contaminant, roughly at the $10 - 30\%$ level at $10$ arcminutes (lower panel of Fig. \ref{ratio}).

There are several issues that are worth exploring in the future.

First, we advocate the measurement of density-ellipticity correlation 
from current weak-lensing surveys. This can be straightforwardly implemented
using current data and increases their scientific return. 
Most surveys are sufficiently deep that
the intrinsic contribution should be small (a few percent; as long as 
$z_{\rm T}$ is sufficiently large -- see \S \ref{nongaussDE}). The same conclusion
was reached regarding the ellipticity-ellipticity correlation by
MWK and CNPT. It would be useful to check that this is indeed the case, 
by measuring both correlations, and see if they are consistent with each other
under the lensing interpretation. Such a check requires knowledge of the galaxy bias $b$ 
as well as the luminosity function (slope $s$), however. The luminosity function is in principle 
directly measurable, while the galaxy bias $b$ can be obtained from higher order clustering
measurements (see Fry 1994, Scoccimarro et al. 2001 and references therein). What makes this
program challenging is that both $s$ and $b$ can be redshift-dependent.

Second, it is important to check for the possible contamination of low 
redshift galaxy-galaxy lensing measurement by
intrinsic alignment. Better measurements of intrinsic alignment in surveys of closeby galaxies
will be quite interesting (e.g. Brown et al. 2000). The SDSS survey is in principle very 
useful for this purpose (McKay et al. 2002, unpublished preprint). 
The different scalings with redshift between the intrinsic and extrinsic signals
can also be used to tell them apart (see eq. [\ref{nongaussDEin2}] \& [\ref{DEex2}]).

It would also be important to check the predictions we make in this paper,
for both ellipticity-ellipticity and density-ellipticity correlations, against
numerical simulations, especially on large scales (above several Mpc/h). 
A non-zero measurement of the density-ellipticity correlation on large scales from simulations
might be interpreted as indicating that non-Gaussian tidal fluctuations are indeed important, 
since Gaussian theories predict a vanishing contribution.
We should emphasize that the normalization of the
correlations is really not predicted by current analytic theories.
It involves free parameters, such as the moments of inertia, which have to be fixed
by matching simulations and/or observations.
Presumably, such a normalization procedure can approximately account for the reduction of
alignment correlations seen in simulations, due to nonlinear effects (e.g. Porciani et al. 2001a,b),
or hydrodynamic effects (e.g. van den Bosch et al. 2002). 
But it is also quite possible that a more radical change in alignment theories is required; testing
this will be important for progress. It is for this reason that our estimates, such as
the level of contamination to galaxy-galaxy lensing due to intrinsic alignment, should be
viewed with some skepticism. 
If the basic picture adopted in this paper holds up, 
the prediction for a large scale linear scaling of the intrinsic correlations with 
the mass power spectrum should be fairly robust. This has important implications, in that
the ratio of the large scale intrinsic to extrinsic signals does not drop significantly with scale. 


We thank Ue-Li Pen for useful comments and Roman Scoccimarro for helpful discussions. 
Support for this work is provided by an Outstanding Junior Investigator Award from the 
DOE and an AST-0098437 grant from the NSF.

\section*{Appendix: \\Estimators for the Density-Ellipticity Correlation in
Real and Fourier Space}
\label{appendix}

An estimator for the real-space density-ellipticity correlation is
\begin{eqnarray}
\hat \xi_{\rm cross} (\Delta\theta)
\equiv {\sum_{\alpha,\beta} (n(\alpha) - \bar n) \epsilon_t
(\beta) n(\beta) \Theta_{\alpha,\beta}^{\Delta\theta} \over
\sum_{\alpha',\beta'} \Theta_{\alpha',\beta'}^{\Delta\theta} \bar n^2}
\end{eqnarray}
where one can imagine pixelizing the survey so that each pixel
contains either one or no galaxy, symbolized by $n(\alpha)$ which
equals $1$ or $0$ depending on whether the pixel $\alpha$ contains
a galaxy or not; $\bar n$ is the average number of galaxies per
pixel, and $\Theta_{\alpha,\beta}^{\Delta\theta}$ is equal to $1$
if pixel $\alpha$ and pixel $\beta$ are separated by $\Delta\theta$, 
zero otherwise. This estimator on the average provides a measure
of $\xi_{\rm cross}$ in eq. (\ref{xicrossIN}). 
The pixelization discussed above can be thought of as a 
conceptual device. In practice, one way to carry out the estimation
is: take all pairs of galaxies at
a given separation of interest $\Delta \theta$, compute the
average $\epsilon_t$, and then do the same for a pair consisting
of a galaxy and a random point; subtracting the two and multiplying
the result by appropriate constants will yield $\hat \xi_{\rm cross}$. 

Another method to estimate $\xi_{\rm cross} (\Delta \theta)$ 
makes explicit use of pixelization. Suppose
the survey is pixelized so that each pixel contains at least
several galaxies. Let $N(\alpha)$ be the number of galaxies in pixel $\alpha$.
We define a pixel ellipticity by
\begin{equation}
\hat \epsilon_t (\beta) \equiv \sum_{i=1}^{N(\beta)} \epsilon_t(i) / \bar N
\label{epsilonthat}
\end{equation}
where $\bar N$ is the average number of galaxies per pixel, and
$i$ labels individual galaxies within the pixel $\beta$. 
Note that $\epsilon_t$ at pixel $\beta$ is defined with respect to
another pixel $\alpha$ (in order to define the tangential part). 

Then, $\xi_{\rm cross}$ can be estimated by:
\begin{eqnarray}
\label{xicrosshat2}
\hat \xi_{\rm cross} (\Delta \theta)
= { \sum_{\alpha,\beta} (N(\alpha) - \bar N) \hat \epsilon_t (\beta) 
\Theta^{\Delta\theta}_{\alpha,\beta} \over 
\sum_{\alpha', \beta'} \Theta^{\Delta\theta}_{\alpha',\beta'} \bar N^2 }
\end{eqnarray}
This estimator would also yield $\xi_{\rm cross}$ (eq. [\ref{xicrossIN}])
on the average.

The reader might wonder why we had not used a seemingly more
natural definition of pixel ellipticity: 
$\hat \epsilon_t (\beta) \equiv \sum_{i=1}^{N(\beta)} \epsilon_t(i) / N(\beta)$.
A moment's thought would reveal that using this instead in 
eq. (\ref{xicrosshat2}) gives 
\\
$\langle {\int d\chi_g W_g (\chi_g) \int d\chi_g' W_g (\chi_g') 
\delta_g (\bi{\theta}, \chi_g) (1 + \delta_g (\bi{\theta'}, \chi_g'))
\epsilon_t (\bi{\theta'}, \chi_g') \over
\int d\chi_g'' W_g (\chi_g'') (1 + \delta_g (\bi{\theta'}, \chi_g''))}
\rangle $
\\
rather than eq. (\ref{xicrossIN}). The above is a legitimate
quantity to consider, but the expectation value then no longer
simply consists of two terms as in eq. (\ref{xicrossIN}). Rather,
there will be an infinite number of terms (see Bernardeau 1998 and
Hui \& Gaztanaga 1999 for discussions of related issues). 

In this paper, we have focused exclusively on the density-ellipticity correlation
in real space. Let us provide here the corresponding estimator
in Fourier space. A Fourier space estimator has the usual advantage of
uncorrelated band-power errors, the complication of window function aside.
We will continue to think of a pixelized survey where each pixel
has at least several galaxies. 

Let us define $\hat \delta(\alpha) = (N(\alpha) - \bar N)/\bar N$ and
its Fourier transform:
\begin{eqnarray}
\tilde \delta (\bi{\ell}) = d^2 \theta \sum_\alpha \hat \delta (\alpha) e^{-i 
\bi{\ell} \cdot \bi{\theta_\alpha}}
\end{eqnarray}
where $d^2 \theta$ is the angular size of each pixel. 

Let us also define the Fourier transform:
\begin{eqnarray}
\tilde \epsilon_j (\bi{\ell}) = d^2 \theta \sum_\beta \hat \epsilon_j (\beta)
e^{-i \bi{\ell} \cdot \bi{\theta_\beta}}
\end{eqnarray}
where $j = 1, 2$, and 
$\hat \epsilon_j (\beta) \equiv \sum_{i=1}^{N(\beta)} \epsilon_j (i) / \bar N$. 
Define also $\tilde \epsilon_t (\bi{\ell}) 
= [(\ell_1^2 - \ell_2^2) / \ell^2] \tilde \epsilon_1 (\bi{\ell})
+ [2 \ell_1 \ell_2 / \ell^2] \tilde \epsilon_2 (\bi{\ell})$ (note that
$\tilde \epsilon_t$ is not the Fourier transform of $\hat \epsilon_t$). 

Finally, we can form the following estimator:
\begin{eqnarray}
\hat P_{\rm cross} (\ell) \equiv {1\over V} 
{\sum_{\bi\ell} \tilde \delta (\bi{\ell}) \tilde
\epsilon_t (\bi{\ell}) \over \sum_{\bi{\ell}}}
\end{eqnarray}
where $V$ is the area of the survey, and $\sum_{\bi\ell}$ refers to
summation over all modes with $|\bi{\ell} | = \ell$. 
Since the intrinsic density-ellipticity correlation vanishes,
this would on the average gives us an estimate of something
related to the extrinsic density-ellipticity correlation in
Fourier space. It can be shown that, ignoring source clustering
corrections,
\begin{eqnarray}
\langle \hat P_{\rm cross} (\ell) \rangle = 
2 \int {d\chi_L \over a^2} 
{\tilde W_L (\chi_L)  \tilde W_L^s(\chi_L)\over r(\chi_L)^2} 
P(k = \ell / r(\chi_L)) \\ \nonumber 
+ 2 \int {d\chi_L \over a} {\tilde W_L^b (\chi_L) W_g (\chi_g = \chi_L) 
\over r(\chi_L)^2} P(k = \ell/r(\chi_L))
\end{eqnarray}
The two terms on the right hand side are Fourier analogs
of the $A$ and $B$ terms in eq. (\ref{exdeltagepsilon}).

             
       
\end{document}